\documentclass[letterpaper,twocolumn,10pt]{article}
\usepackage{usenix2019,epsfig,endnotes}

\usepackage{mathptmx} 

\newcommand{\ignore}[1]{}
\usepackage{fancyhdr}
\usepackage[normalem]{ulem}
\usepackage[hyphens]{url}
\usepackage{microtype}

\usepackage{enumerate}
\usepackage{graphicx}
\usepackage{listings}
\usepackage{tabularx}
\usepackage{amsmath}
\usepackage{color}
\usepackage{amssymb}

\usepackage{mathtools}
\usepackage{makecell}
\usepackage{subfig}
\usepackage{caption}
\usepackage{xcolor}
\usepackage{booktabs}
\usepackage{array}
\usepackage{algorithm,algpseudocode}
\usepackage{multicol}
\usepackage{pifont}
\usepackage{nicefrac}
\usepackage{xspace}
\usepackage{authblk}

\algtext*{EndWhile}
\algtext*{EndIf}
\algtext*{EndProcedure}

\usepackage{cite}

\newcommand{\RNum}[1]{\uppercase\expandafter{\romannumeral #1\relax}}

\newcommand{\red}[1]{\textcolor{black}{#1}}
\newcommand{\green}[1]{\textcolor{black}{#1}}

\newcommand{\anvm}{{AsymNVM}\xspace}

\fancypagestyle{firstpage}{
  \fancyhf{}

  \fancyhead[C]{} 
  \fancyfoot[C]{\thepage}
}  

\title{A Case for Asymmetric Non-Volatile Memory Architecture} 
\author[1]{Teng Ma}
\author[1,3]{Mingxing Zhang}
\author[1]{Kang Chen}
\author[2]{Xuehai Qian}
\author[1]{Yongwei Wu}
\affil[1]{Tsinghua University}
\affil[2]{University of Southern California}
\affil[3]{Sangfor Inc}

\begin{document}
\maketitle
\thispagestyle{firstpage}
\pagestyle{plain}

\begin{abstract}

The byte-addressable Non-Volatile Memory (NVM) is a promising technology since it simultaneously provides DRAM-like performance, disk-like capacity, and persistency.
The current NVM deployment is {\em symmetric},
where NVM devices are directly attached to servers.
Due to the higher density, NVM provides much larger
capacity and should be shared among servers. 
Unfortunately, in the symmetric setting, 
the availability of NVM devices is affected 
by the specific machine it is attached to. 
High availability can be realized by replicating 
data to NVM on a remote machine.
However, it requires full replication of data 
structure in local memory, limiting the size of the working 
set.

This paper rethinks NVM deployment and 
makes a case for the {\em asymmetric} non-volatile
memory architecture, 
which decouples servers from persistent data storage.
In the proposed {\em \anvm architecture},
NVM devices (i.e., back-end nodes)
can be shared by multiple servers (i.e., front-end nodes)
and provide recoverable persistent data structures.
The asymmetric architecture is made possible by 
the high-performance network (e.g., RDMA), and 
follows the recent industry trend of 
resource disaggregation. 

We build {\em \anvm framework} based on 
\anvm architecture that implements:
1) high performance persistent data structure update;
2) NVM data management;
3) concurrency control; and 
4) crash-consistency and replication. 
The central idea is to use {\em operation logs}
to reduce the stall due to RDMA writes and 
enable efficient batching and caching in front-end
nodes. 
To evaluation performance,
we construct eight widely used data structures and 
two applications based on \anvm framework, and
use traces of industry workloads.
In a cluster with ten machines (at most seven machines
to emulate a 60GB NVM using DRAM with additional latency),
the results show that \anvm achieves comparable
(sometimes better)
performance to the best possible symmetric architecture while avoiding
all the drawbacks with disaggregation.
Compared to the baseline \anvm, speedup brought by the proposed optimizations is drastic, 
--- 6$\sim$22$\times$ among all benchmarks.



\end{abstract}

\section{Introduction}\label{sec:intro}

Emerging Non-Volatile Memory (NVM) is 
blurring the line between memory and storage.
These kinds of memories, such as 3D X-Point~\cite{3dxpoint}, phase change memory (PCM)~\cite{burr2010phase, lee2009architecting, zhou2009a}, 
spin-transfer torque magnetic memories (STTMs)~\cite{apalkov2013spin-transfer}, and memristors 
are byte-addressable, and provide
DRAM-like performance, high density, and persistency
at the same time.
To unleash the potential of NVM,
the existing solutions attach 
NVM directly to processors~\cite{giles2015softwrap, coburn2011nv, volos2011mnemosyne, liu2017dudetm, venkataraman2011consistent}, enabling
high-performance implementations of persistent data structures using load and store 
instructions on local memory.

While accessing NVM via local memory provides
good performance, 
it may not be the most suitable setting 
in the data center due to the 
{\em desire of sharing NVM}.
Due to the higher density, NVM can provide much larger capacity~\cite{mittal2016asurvey},
which would likely exceed the need of a single machine~\cite{arulraj2017build}. 
Moreover, in data center servers,
the resources are often under utilized~\cite{eisenman2018reducing, delimitrou2014quasar},
--- Google's study~\cite{googledata} shows 
the resource utilization lower than 40\% on average.
We expect that persistent resource utilization
will follow the same trend. 





To enable NVM sharing, recent work~\cite{shan2017distributed}
builds a distributed shared persistent memory system.
The system provides a global, shared, and persistent 
memory space for
a pool of machines with NVMs attached to each 
at the main memory bus.
Once an NVM device is attached to a specific machine, its data become unavailable
when the host machine goes down.
It violates the availability requirement of 
many data center applications.
To ensure {\em availability},
one still needs to replicate the data to a remote NVM~\cite{islam2016high}.
However, it requires full replication 
of data structures in local memory, 
limiting the size of the working set. 



In essence, these challenges are due to the 
{\em symmetric} nature of the current NVM
deployment~\cite{islam2016high, stuedi2017crail, lu2017octopus}.
To fundamentally overcome the drawbacks, 
we rethink NVM deployment and make
a case for the {\em asymmetric} NVM architecture, 
in which NVM devices are not 
associated with the individual machine and 
accessed only {\em passively via fast network}.
In this {\em \anvm} architecture, the number of NVM devices can be much smaller than the number of machines,
and they can be provided as specialized ``blades''.

\anvm architecture is an example of the recent
trend of disaggregation architecture,
which was first proposed for capacity expansion and memory resource sharing by Lim {\em et al.}~\cite{lim2009disaggregated, lim2012system}. 
As described by Gao {\em et al.}~\cite{gao2016network}, 
disaggregated architecture is a paradigm
shift from servers each tightly integrated 
with a small amount of
various resources (i.e., CPU, memory, storage)
to the disaggregated data center built as a pool of
standalone resource blades and interconnected using a network fabric. 
In industry, Intel's RSD~\cite{intelwhitepper} and HP's ``The Machine''~\cite{hpdc} are state-of-the-art disaggregation architecture products.
Such systems are not limited by the 
capability of a single machine and can provide
better resource utilization and the 
ease of hardware deployment. 
Due to these advantages, 
it is considered to be the future of data centers~\cite{intelrackdc, intelrsddc, inteldisaggregation, hpdc, hsu2018smoothoperator}.
In \anvm architecture, the NVM devices are 
instances of disaggregated resources that are 
not associated with any server.
It also improves the availability 
since the crash of a server will not affect
NVM devices. 



\anvm architecture (and resource disaggregation) relies on the high-performance network, e.g., RDMA,  
which provides the direct remote access 
capability with RDMA verbs 
(\texttt{RDMA\_Write}, \texttt{RDMA\_Read}, etc.), as well as reliable connection.
While promising, the architecture poses 
key challenges. 
First, 
a straightforward implementation of replacing the is local store/load instructions  
with \texttt{RDMA\_Write} and \texttt{RDMA\_Read} operations suffers from long network latency. 
Although the throughput of RDMA over InfiniBand is comparable 
to the throughput of NVM, the NIC 
cannot provide enough 
IOPS for fine-grained data structure accesses.
Second, the management of smaller local 
volatile memory needs to be cleared designed
to ensure high performance. 
Third, the interface of back-end nodes
needs to be simple enough to ensure reliability while
ensuring efficiency.


Based on \anvm architecture, 
this paper builds {\em \anvm framework}
for implementing 
data structures and applications with 
high performance and availability.
The framework efficiently implements four
components. 
1) {\em high performance persistent
data structure update} realized by 
introducing {\em operation log} to 
reduce the stall due to RDMA writes and enable
efficient batching and caching; 
2) {\em NVM data management} that handles 
non-volatile memory allocation and free, and
metadata storage;
3) {\em concurrency control} for lock-free
and lock-based data structures; and
4) {\em crash-consistency and replication}
that ensures correct recovery and availability.
Putting all together, \anvm framework
efficiently solves the three challenges
discussed before. 

To evaluate the performance of 
\anvm framework, 
we construct eight widely used data structures
(i.e., stack, queue, hash-table, skip-list, binary search tree, B+tree, multi-version binary search tree, and multi-version b+tree) and
two applications (i.e., TATP and SmallBank) 
based on \anvm framework, and
use traces of industry workloads.
The data structures and applications are executed
in a cluster with ten machines, 
among them at most seven machines
to emulate a 60GB NVM using DRAM with additional latency.
The results show that \anvm achieves comparable
(sometimes better)
performance to the best possible 
symmetric architecture while avoiding
all the drawbacks with disaggregation.
Compared to the baseline \anvm, speedup brought by the proposed optimizations is drastic, 
--- 6$\sim$22$\times$ among all benchmarks.

The remainder of this paper is organized as follows.
Section~\ref{sec:back} discusses the
current deployment of the NVM devices.
Section~\ref{sec:overview} presents the overview
of \anvm architecture and framework.
Section~\ref{sec:persist} explains the mechanisms
to ensure efficient persistent updates. 
Section~\ref{sec:mm} discusses the NVM data
management.
Section~\ref{sec:concurrency} and 
Section~\ref{sec:recovery} explains details of concurrency
control, recovery, and replication. 
Section~\ref{sec:eva} evaluates \anvm architecture
and framework. 
Section~\ref{sec:rel} discusses additional 
related work. 
Section~\ref{sec:con} concludes the paper. \looseness=-1

\section{Background and Motivation}\label{sec:back}

We consider two current architectures 
of deploying NVM devices:
1) single-node setting that only considers one machine and one NVM device;
2) symmetric distributed setting, where each machine in the cluster is attached with an NVM device.
\subsection{Single-Node Local NVM}

To leverage the advantages of
DRAM-like performance of 
byte addressable NVM, 
recent studies consider the setting
that NVM device is directly accessed via 
the processor-memory bus using load/store instructions.
This design avoids the overhead of legacy block-oriented file systems or databases. 
Persistent memory also allows programmers to update persistent data structures directly 
at byte level without the need for serialization to storage.

Based on this setting, many kinds of persistent data structures are proposed.
For example,
CDDS-Tree~\cite{venkataraman2011consistent} uses multi-version to support atomic updates 
without logging.
NV-Tree~\cite{yang2015nv} is a consistent and cache-optimized B+Tree, which reduces CPU cache line flush operations.
HiKV~\cite{xia2017hikv} constructs a hybrid index strategy to build a persistent key-value store.
Since all the data accesses are performed by local store/load instructions,
these implementations can offer the best performance.
Although these persistent data structures can survive a failure of the machine,
they are not accessible during the recovery/restarting. 

\subsection{Symmetric Distributed NVM}

Symmetric architecture is widely used in distributed systems (e.g., shared memory and distributed file systems). In symmetric architecture, each machine has its own NVM device. 
To achieve good availability on top 
of persistency, 
one needs to replicate its data structures 
to multiple NVM devices.
Mojim~\cite{islam2016high} implements this mechanism by adding two more synchronization APIs 
(\texttt{msync/gmsync}) in the Linux kernel.
Specifically, it allows users to set up a pair of primary node and the mirror node.
Once these synchronization APIs are invoked,
Mojim efficiently replicates fine-grained data from the
primary node to the mirror node using an optimized RDMA-based protocol.
This synchronization is implemented by appending 
primary node's logs with end marks to 
the mirror node's log buffer, 
thereby tolerating a failure of the primary node.
Mojim also allows users to set up several backup nodes that are the only
weakly-consistent replication of data
in the primary node.

With a similar interface, Hotpot~\cite{shan2017distributed} extends Mojim to a distributed shared persistent memory system.
It provides a global, shared, and persistent memory space for
a pool of machines with NVMs attached at the main memory bus.
As a result, applications can perform native memory load and store instructions to
access both local and remote data in this global memory space, and
can at the same time make their data persistent and reliable.
To achieve this, when a committed
page is written, Hotpot creates a 
local copy-on-write page and marks it as dirty.
These pages are not write-protected
until they become committed after an explicit invoking of the synchronization APIs.
At this point, the modifications
to this page from all nodes are finished.

The two systems are designed for the 
symmetric usage of NVM.
As a result, Mojim requires a full replication of the data structure in local memory.
Similarly, Hotpot assumes that the dirty page can always be held in memory 
and only uses a simple LRU-like mechanism to evict redundantly and committed pages.
\section{\anvm Overview}
\label{sec:overview}

\subsection{\anvm Architecture}
\label{sec:arch_overview}

To overcome the drawbacks of 
the current single-node or symmetric 
distributed NVM architecture, 
we make a case for the asymmetric non-volatile
memory architecture, i.e., {\em \anvm architecture}.
In this asymmetric distributed setting, 
the number of NVM devices 
can be much smaller than the number of machines, 
and they can be even attached to only a few 
specialized ``blades''. 
Thus, NVM device/blade is shared by multiple client machines,
and the memory space of these client machines may be much smaller than the capacity of the NVM devices.

\begin{figure}[!t]
  \centerline{\includegraphics[width=0.9\columnwidth]{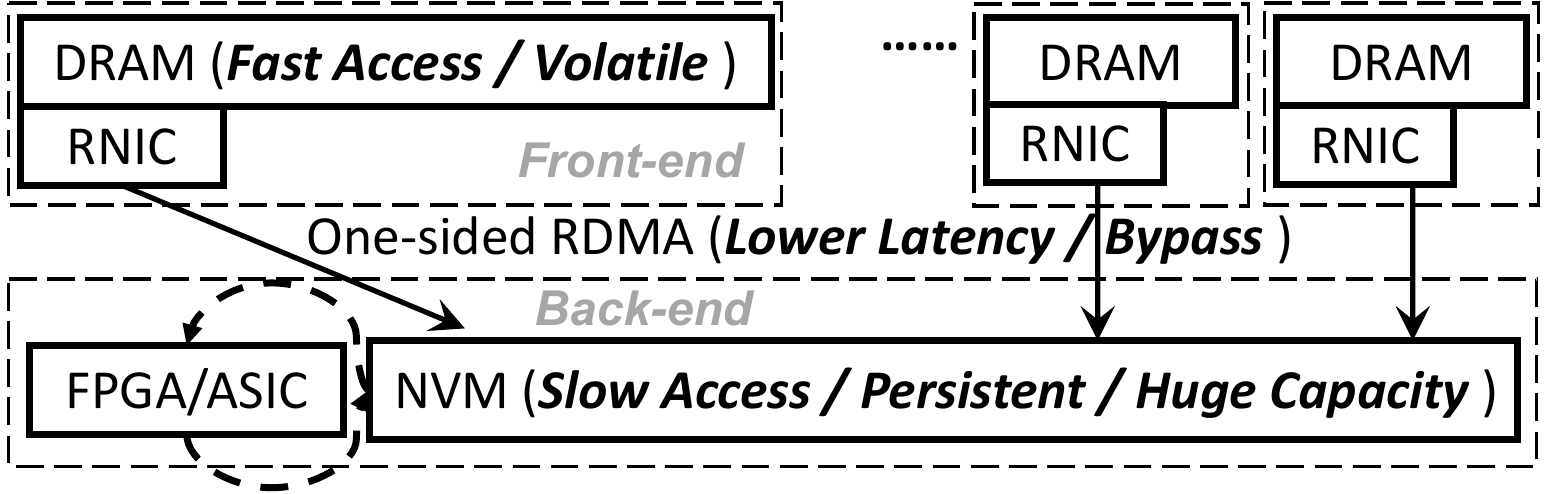}}
\caption{\emph{\anvm} Architecture}
\label{fig:overall}
\end{figure}

Figure~\ref{fig:overall} shows the 
\anvm architecture,
which includes two components performing
different functions:
1) {\em back-end nodes} that have NVM attached to their memory bus;
and 2) {\em front-end nodes} that actually operate the data structures on NVM.
In \anvm architecture,
front-end nodes can only access 
back-end nodes via fast network. 
Specifically, the relationship between 
front-end and back-end nodes is ``{\em many-to-many}''
(i.e., a front-end node can access multiple back-end nodes and a back-end node can also be shared by multiple front-end nodes).
Compared to the symmetric architecture, 
\anvm architecture offers three advantages:
a) it follows the promising trend of 
disaggregated data centers;
b) it naturally matches the desire of 
sharing NVM devices; 
c) it can ensure availability with
multiple back-end nodes; and 
d) the potentially simpler back-end nodes 
leads to better reliability~\cite{lampson1983hints}.

To the best of our knowledge, there are very 
few works on asymmetric distributed NVM. Octopus~\cite{lu2017octopus} can support asymmetric service with file interface but its current implementation is still using the symmetric setting.
Mitsume~\cite{tsai2018mitsume} is the only work that can naturally support asymmetric deployment of NVM devices for providing object store interface. We make a step further to support {\em byte addressable data structures}.

\subsection{Key Challenges}
\label{sec:arch_challenge}

Despite the benefits, 
\anvm architecture also brings three key challenges.

The first challenge is {\em network latency}.
Although the bandwidth of InfiniBand is 
comparable to NVM, 
the RDMA operation RTT is about 2 $\mu s$, 
which is much larger than the latency of NVM write (about 200 $ns$).
If we simply replace local read/write operations with \texttt{RDMA\_Read} and \texttt{RDMA\_Write} 
operations, 
the performance will significantly degrade.

The second challenge is how to 
efficiently use {\em the small volatile space} of the front-end nodes.
Keeping a full copy of the data structure 
in the front-end (like Mojim) can always offer
the best performance.
However, it contradicts with the original 
purpose of the asymmetric setting.
The high density of NVM devices 
makes it capable to hold terabytes of data.
On the other side, the memory space of the front-end is typically to be
only tens of gigabytes.
This asymmetry means that 
only the necessary data in the current work-set 
should be loaded into the front-end for caching.

The third challenge is how to design 
the interface of the back-end nodes 
that is {\em both simple and effective}.
The most straightforward interface is 
to assume that the back-end nodes 
can be programmed to perform arbitrary
RPC calls.
However, to ensure high
reliability, the back-end nodes need to be
simple.
Therefore, they 
should be only asked to perform 
a small collection of simple APIs,
such as remote memory read/write, remote memory allocation/release, lock acquire/release, etc.
With a limited interface that constitutes 
a small set of {\em fixed} APIs, it is possible to
implement the {\em simple} logic of the back-end nodes in specialized hardware 
(e.g., using ASIC or FPGA), 
instead of using a general-purpose CPU.

\subsection{\anvm Framework Overview}
\label{sec:framework_overview}

To address the challenges, 
we build the {\em \anvm framework}, 
a general framework for 
implementing high-performance data structures 
based on \anvm architecture.
The \anvm framework assumes that all 
persistent data are hosted on the NVM devices 
in the remote back-end nodes,
and can be much \emph{larger} than the limited size of the local volatile memory in the front-end nodes.
Moreover, the back-end nodes are always {\em passive}:
they never actively start a communication 
with the front-end nodes,
but only passively response to the API invocations from the front-end.

We assume the back-end nodes are equipped with advanced NIC that supports RDMA. 
In other words, the front-end nodes can 
directly access their data
via \texttt{RDMA\_Read}/\texttt{Write} operations.
Although it is possible to implement any kind of data structures with only RDMA verbs,
the performance will suffer. 
In addition, the back-end nodes need to expose
methods to allow front-end nodes to manage
NVM data. 
To this end, the \anvm framework implements
two sets of \emph{simple and fixed} API functions
in the back-end, \red{beyond} RDMA verbs. 

The first set of APIs provides  
a {\em transactional interface}
that allows the front-end nodes to push a list of 
update logs to the back-end for persistency,
which is guaranteed to be executed in an all-or-nothing manner.
The transactional interface is simple and has
two variants.
Specifically, a transaction can include:
1) a collection of {\em memory logs}, 
--- \{memory address, value\} pairs; or
2) an {\em operation log}, which includes
the operation and parameters applied to a
certain data structure. 
The operation log is used to reduce the stall
due to remote persistency, and more details
will be discussed in Section~\ref{sec:persist}.
The back-end nodes ensure 
that all these addresses are updated atomically. 

The second set of APIs handles {\em memory management}, which  
includes remote memory allocation, releasing, and global naming.
The \anvm framework implements a two-tier slab-based memory allocator~\cite{bonwick1994slab}. The back-end runs in the remote NVM to ensure persistency and provide the fixed-size blocks. The front-end deals with the finer-grained memory allocations. 
Section~\ref{sec:mm} is dedicated to 
NVM data management.

The \anvm framework supports SWMR (Single Writer Multiple Reader) access model by concurrent control mechanisms.
This means that if two front-end nodes
perform writes on the same address, 
they should be synchronized by locks. 
In addition, the framework assumes that reads and 
writes to the same address are also properly
synchronized by locks. 
Based on different applications, we can support lock-free and lock based data structures.
We do not implement the API
for concurrency control, 
e.g., to implement certain lock mechanism,
instead we leverage existing
RDMA primitives.
The details are discussed in Section~\ref{sec:concurrency}.

Finally, the \anvm framework 
adopts a consensus-based voting system to 
detect machine failures. The 
details about recovery and replication
are discussed in Section~\ref{sec:recovery}.

\subsection{RDMA Operations}

There are two common programming paradigms of RDMA. The straightforward approach is 
the two-sided server-reply~\cite{kalia2014using, jakiro_memcachedrdma} paradigm which directly replaces traditional send and receive with \texttt{RDMA\_Send} and \texttt{RDMA\_Recv}.
The other one is server-bypass~\cite{mitchell2013using, dragojevic2014farm, su2017rfp} paradigm using one-sided RDMA, which requires the 
system re-design to exploit such feature~\cite{mitchell2013using, dragojevic2014farm, su2017rfp}.
The \anvm framework uses one-sided RDMA to improve the performance.
Besides, it use polling 
to detect completions similar to~\cite{zhang2015mojim}. 
This means that the front-end nodes can access the memory space on remote NVM devices
directly via \texttt{RDMA\_Write}, \texttt{RDMA\_Read}, and even RDMA atomic operations
without notifying the processing unit 
(e.g., CPU or FPGA/ASIC chips) on the remote side.

In \anvm architecture, back-end nodes need to manage metadata consistently. RDMA provides several atomic verbs to guarantee that any update to a {\em 64-bit} data is atomic. Thus, we can 
apply RDMA atomic operations to the critical metadata, e.g., root pointer of data structure.

Due to the non-volatile nature of the remote NVM, 
the data may be corrupted if the 
back-end crashes during a single \texttt{RDMA\_Write} operation. \anvm framework 
guarantees the data
integrity via checksum.

\section{Efficient Persistent Update}
\label{sec:persist}

\subsection{Basic Implementation}

At low level, the persistent data structure implementation should support read/write 
operations. 
A {\em read} can return data that is not yet persisted,
but if there is a persistent \green{fence~\cite{pelley2014memory, kolli2016delegated}},
the read should return the persisted data
if it is produced before the fence. 
When a {\em write} (update) returns, the data should always be persisted
in the back-end NVM.
The straightforward implementation of the two 
operations is to perform 
\texttt{RDMA\_read} and \texttt{RDMA\_write}
on the back-end nodes.


However, the simple implementation may incur 
considerable rounds of network communications.
As we discussed earlier,
the network latency is much higher than memory writes,
thus the performance will suffer.
In addition, 
the volatile local memory 
is typically much smaller than the whole data structure in remote NVM,
so the front-end nodes need to use proper data
eviction mechanisms when using DRAM as the cache. 


\subsection{Decoupled Memory Log Persistency}

To reduce the performance impact of persistency,
DuduTM~\cite{liu2017dudetm} uses redo log and decouples 
the update of real data structure in NVM and 
the persistency of redo log.
In another word, a write can return after the 
redo log is persisted and does not need to 
wait for the data structure modification.
In \anvm framework, we also use memory log to 
improve performance. 
Unlike prior works, in \anvm architecture, 
the front-end and back-end nodes are 
distributed, the only reasonable choice
is to use redo log. 

In \anvm framework, each write (update) will 
generate a number of {\em memory logs},
and the back-end node provides the 
transaction APIs to ensure that the memory logs
are persisted atomically and in an all-or-nothing
manner. 
When memory logs in the transaction is 
persisted, the back-end node sends back
an acknowledgement, so that the write in the 
front-end can return and is guaranteed to be durable.
The back-end node also guarantees that 
the modifications to the real data structure 
are performed (i.e., {\em replaying} the persisted logs) in order 
due to the sequential log writing. 

\begin{figure}[!h]
  \centerline{\includegraphics[width=0.9\columnwidth]{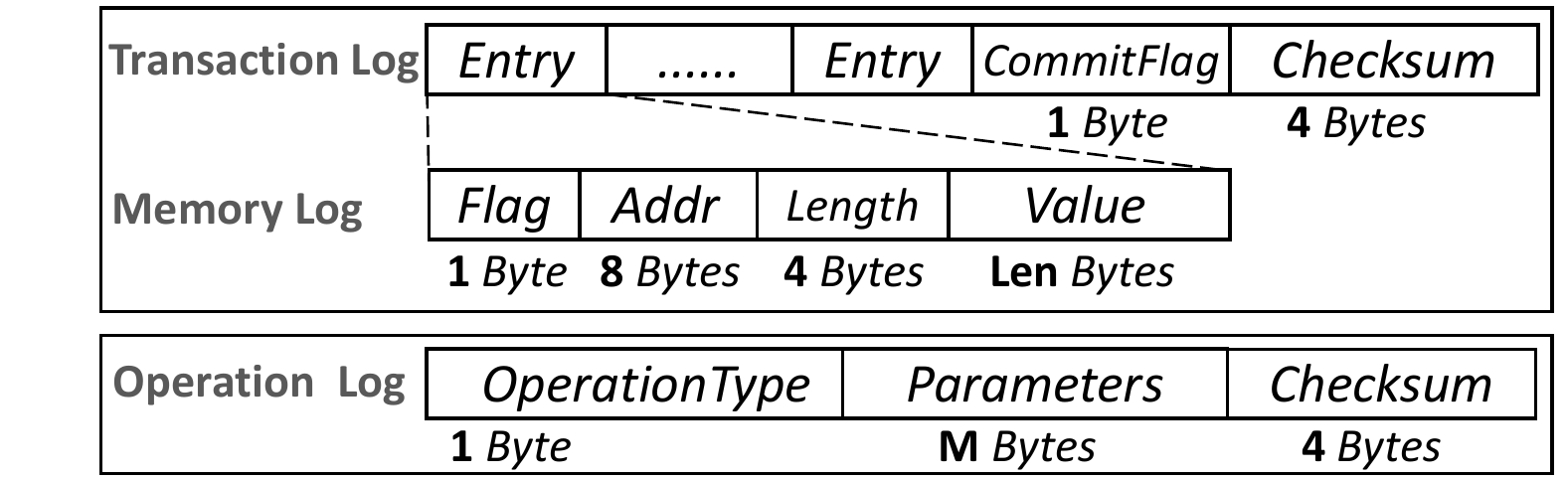}}
\caption{Memory Log vs. Operation Log}
\label{fig:format}
\end{figure}

Specifically, the transaction API is \texttt{remote\_tx\_write}.
The input parameter is a list of \{address, value\} pairs, each of which consists of a memory address and a value that should be written to this address.
The back-end nodes keep two areas:
the {\em data} area holds the real data structures; 
the {\em log} area records the transaction logs.
The front-end can directly read the data area,
but any updates have to go through the log area.
To implement \texttt{remote\_tx\_write},
the framework library will construct a continuous set of memory logs and append to the corresponding log area in remote NVM via a single \texttt{RDMA\_Write} operation.
The format of these memory logs is shown in
Figure \ref{fig:format}.
Every log entry includes \textit{address, length, data}, and \textit{one-byte flag} in the head. This flag indicates whether the value is in the memory log, it is used by an optimization related to batching, more details will be discussed in   Section~\ref{sec:batching}. 
A transaction will produce several log entries, a commit flag, and a \emph{checksum} value.
The checksum of a transaction 
is recorded as the end mark and can be 
used to validate the integrity of the appended log.
After the restart of the back-end node, it needs to use the checksum of the last transaction to validate whether all the log entries of this transaction is flushed to the NVM.

The advantages of using the transactional API are
1) reducing the persistency latency due to 
modification of real data structure; and 
2) largely reducing the required rounds of RDMA operations.
Without the transaction API, multiple rounds of RDMA operations are needed when writing to
multiple non-continuous areas of the NVM, or
a continuous area with the size larger than a cache line.
Other works~\cite{remoteaccess, rdmawithpmem, rdmaandnvm} propose to add an additional flush operation to the RDMA standard.
However, such resolution will at least add the additional latency of invoking this flush operation.
Moreover, the additional operation itself does not make the other RDMA operation crash-consistent.
Importantly, the implementation of the transactional API is fixed and simple, 
improving the reliability of back-end nodes.

\subsection{Batching and Caching with Operation Log}
\label{sec:batching}

To further reduce the latency to data 
persistency, we propose the notion of
{\em operation log}, which is shown
at the bottom of Figure~\ref{fig:format}.
Different from the memory logs, 
each write only incur one operation log, 
which contains operation type, 
parameters, and checksum.
A write can return after the operation log
is persisted in the back-end node. 
Persisting operation log can be achieved by 
a {\em single} one-sided RDMA write to the 
back-end node. 

The crucial benefit of operation log is
that it enables {\em batching and caching}.
Once the operation logs are recorded, 
the modifications on the real data structure
can be postponed and batched to improve the performance while ensuring crash consistency
(e.g., asynchronous execution to remove network latency from the critical path, and
combining redundant writes to reduce write operations). 
This is because, even after a crash, 
the proper state can be restored by replaying 
the operation logs that are not executed (i.e., have not yet modified the data area).

It is important to understand the difference 
between the operation log and memory log. 
With only memory log, we can only realize the 
``postpone'' aspect, --- the real data structure
modification can be delayed as long as the 
memory log is persisted. 
But it cannot achieve the ``batched'' aspect, 
because the memory logs of {\em each write} 
need to be persisted with a \texttt{remote\_tx\_write}.
The operation log achieves batching
by combining the memory logs of multiple
writes into one \texttt{remote\_tx\_write}.
At lower level, operation log reduces the 
number of \texttt{RDMA\_Write} operations.
With memory log only, each write needs
at least two \texttt{RDMA\_Write} operations,
--- one for the commit and the others for at least one memory log. 
A write typically needs more than $1$ memory log, 
thus the number of \texttt{RDMA\_Write} operations is usually larger than two.
With the operation log, each write still needs
one \texttt{RDMA\_Write}, but no commit
is required, because the operation log
already serves the purpose of committing the
operation. 
The number of \texttt{RDMA\_Write} operations for 
the memory logs is less
since multiple writes can be opportunistically
coalesced into one \texttt{RDMA\_Write},
depending on the addresses.
In addition, the commit for the batched 
memory logs (not for each write) also 
needs an \texttt{RDMA\_Write}.

\begin{figure}[!t]
  \centerline{\includegraphics[width=1\columnwidth]{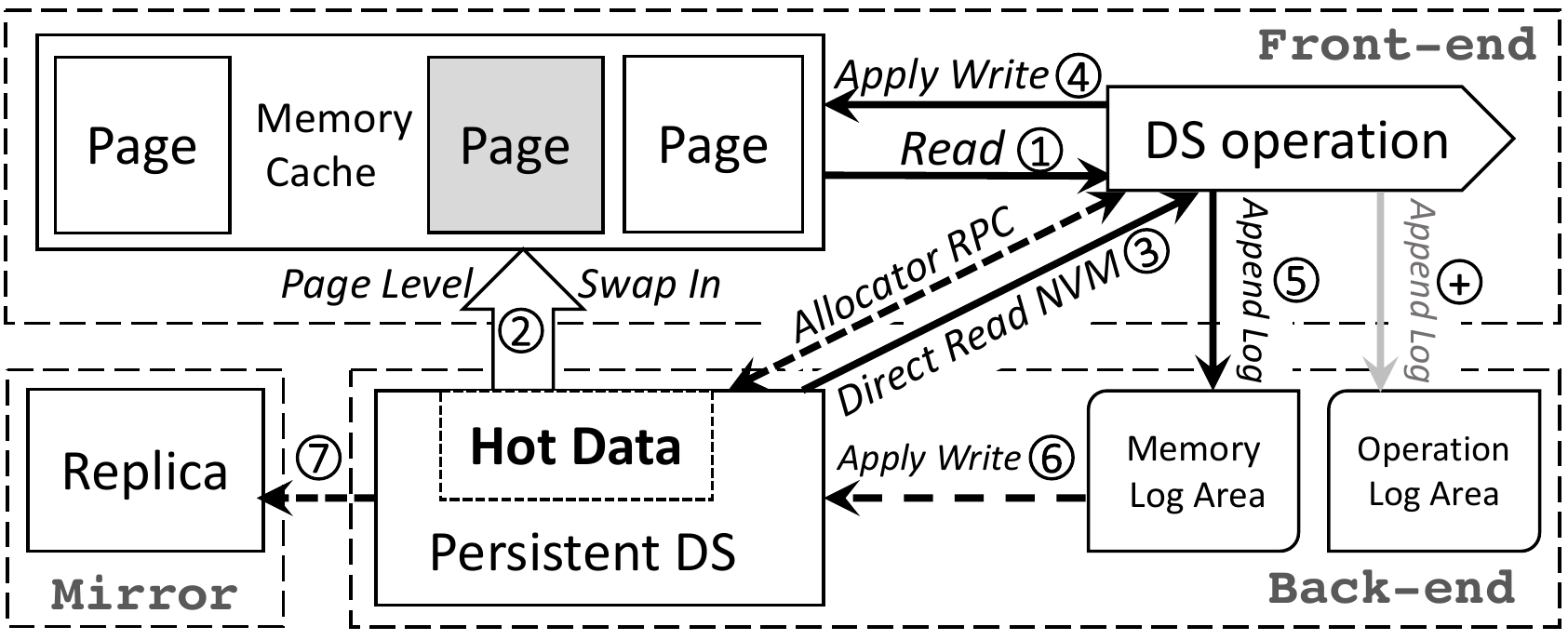}}
\caption{\anvm Framework Data Access Workflow}
\label{fig:dataflow}
\end{figure}

Figure~\ref{fig:dataflow} shows the workflow of 
\anvm framework data accesses. 
Each data structure level 
modification operation typically 
needs to first read the data and then write to 
perform the modifications.
Accordingly,
we divide each operation into two parts 
and use a \emph{Gather-Apply} model: 
gathering data and applying the modifications, and read-only operation only needs the gathered part. 
We use the terms gather and apply, instead of
read and write, to explain the holistic flow
including batching and caching, which 
may involve multiple reads/writes. 
Batching can execute multiple operations together and {\em coalesce} memory logs to both reduce the
number of \texttt{RDMA\_Write} and \texttt{RDMA\_Read} operations. 
Besides, caching will reduce the 
number of \texttt{RDMA\_Read} operations during the gathering phase. They are applicable to all data structures.

\textbf{Gather Data: } The data are fetched from the front-end cache whenever possible (cache-hit, \textcircled{1}). If not cache-hit, 1) the data will either be read from the back-end directly by using \texttt{remote\_nvm\_read} (\textcircled{3}) or, 2) its corresponding page will be swap-in (\textcircled{2}) via \texttt{remote\_nvm\_read} and put to the cache in the front-end memory.
Then, data is read from front-end cache via \textcircled{1}. 
The choice between these two strategies depends on \textit{specific} data structures and follows a principle that using swap-in (\textcircled{2}) for cold data and \texttt{remote\_nvm\_read} (\textcircled{3}) 
for hot data.
Hot data (e.g., the root of a B+Tree) are
accessed frequently than cold data (e.g., the leaf of a B+Tree). 
On a persistent fence, the read after the fence
will need to wait until memory logs before the fence
persisted in the back-end node.

\textbf{Apply Modification: } 
Each modification operation causes one operation log to be flushed to the back-end for recovery (\textcircled{+}). 
The operation log with format \{insert\_op, key, value\} as shown in Figure~\ref{fig:format} will be put in the operation log area. 
Then, the memory logs of format \{address, data\} are be generated afterward. They do not need to be flushed immediately. 
We replace actual data in memory log
with a pointer to the previous flushed operation log to reduce the size of data write,
the data/pointer is indicated by the ``Flag''
of memory log in Figure~\ref{fig:format}.
It is correct because,
after the operation log is stored in the back-end, the data structure modification is persistent and 
recoverable. 
While flushing the logs, 
the cached data (if exist) are modified accordingly (\textcircled{4}). 
If a number of operations get executed successfully, or the buffer is full, the buffered memory logs, together with appended \texttt{TX\_COMMIT}, will also be flushed to the back-end NVM via  \texttt{remote\_tx\_write} (\textcircled{5}). 
These logs are then handled by the back-end (\textcircled{6}) and replicated to the mirror-node (\textcircled{7}) (Section~\ref{sec:recovery} 
discusses details on replication). 
If the back-end fails, the front-end handles exceptions, abort the transaction and clear the cache.

To support a data structure even larger than the capacity of the NVM in a single back-end node, 
\anvm framework supports a distributed data structure partitioning across multiple back-ends. Specifically, the distributed data structure is partitioned with key-hashing. 
When the front-end node executes a data structure operation, it first locates the appropriate back-end according to the key's value using the hash. Then, the front-end will read or modify data using \texttt{remote\_nvm\_read} / \texttt{remote\_tx\_write} in the corresponding back-end node.
At this point, 
the processing is similar to the single back-end scenario.

\subsection{Data Cache in Front-end}\label{subsub:cache}
Several recent works build NVM systems using DRAM as cache~\cite{qureshi2009scalable, liu2017dudetm, xia2017hikv, lee2011energy}. Bw-tree~\cite{levandoski2013bw} uses a cache layer to map logical pages (Bw-tree nodes) to physical pages. We use a similar data structure of hash map to translate the address of data structure nodes in NVM to address in DRAM. Each item in the hash map represents the page cached. The page size is adjustable according to different data structures.

Our cache replacement policy combines the methods of LRU (Least Recently Used) and RR (Random Replacement). LRU works well in choosing hot data,
but its implementation is expensive.
RR is easy to implement but does not provide any guarantee of preserving hot data. 
We use a hybrid approach, ---
first choosing a random set of pages for replacement (RR) and then selecting a least used page from the set to discard (LRU). No page flush is needed because the write workflow already put the memory logs in the back-end node. In a micro-benchmark, the hybrid approach (29.2\%) can reduce the miss ratio by 33.5\% compared to RR (62.7\%) when the size of choosing set is 32, and gain a similar miss ratio as LRU with nearly 27.5\% throughput improvement. 

\subsection{Data Structure Specific Optimizations}
\label{sec:ds_opt}

The \anvm framework are general
to implement different persistent data structures
with high performance. In this section, we 
investigate several commonly-used data structures
and propose data structure specific optimizations.

{\bf Stack/Queue}
We implement Stack and Queue by using the List data structure. Because the only data items that can be accessed in Stack or Queue are headers or tails
and they are more frequently accessed,
the front-end nodes only need to cache nodes pointed by them to reduce \texttt{remote\_nvm\_read}.
If there are not enough data items of headers and tails in the cache i.e., less than a threshold, the front-end will fetch back corresponding data to the cache. 
Moreover, due to the access pattern, 
the operations may be combined because the operations are only allowed on stack header for Stack, and on queue tail for Queue. Thus, the effective pushes will be {\em annulled} by pops for Stacks, and the effective enqueues will be annulled by dequeues for Queue. 
Such an opportunity can be identified by checking
the un-executed operation logs in the front-end
memory. 
For example, for a pop operation to 
the stack data structure, 
we need first to count the number of un-executed push and pop operations in the operation log.
If the number of pushes is larger than the number of pops,
there is no need to access the data area.
This optimization based on operation log
will reduce the RDMA reads and writes.

{\bf Tree-Like Data Structure}
Tree-like data structures have the hierarchical organization. The nodes in higher (near the root) level are more frequently accessed than lower level nodes. Based on this natural property,
we choose to cache higher level nodes with 
higher priority. 
Specifically, the front-end sets a threshold $N$ and the nodes with level larger than $N$ will not be cached. They will be directly accessed through \texttt{RDMA\_Read}. $N$ is dynamically adjusted according to the cache miss ratio $\alpha$, i.e., if $\alpha>50\%$, $N=N-1$ while if $\alpha<25\%$, $N=N+1$. Otherwise, $N$ stays unchanged. The native LRU algorithm treats higher level nodes and lower level nodes in the same way, and hence incurs frequently cache misses. Compared with the primitive LRU algorithm, our mechanism gives a ``hint'' to cache the hot nodes.  

In addition, because trees are sorted data structures, the performance can be improved when the operations are sorted. 
Based on this insight, we pack the sorted operations into a \emph{vector operation}. The operation goes from the root of the tree down to the leaf nodes. The vector can then be split accordingly. The operations in vector segments can be executed in parallel.

\begin{figure}[htb]
  \centering
  \begin{minipage}{0.95\linewidth}
    \begin{algorithm}[H]\footnotesize

\caption{Vector Insert}\label{algo:vector}
\begin{algorithmic}[1]
\Procedure{Vector Insert}{$kvs$}\Comment{keys are sorted}
      \State{$node \gets root$}
      \State{$queue.push(<0, len, node>)$}
      \While{$\text{queue is not empty}$} 
        \State{$begin,end,node \gets queue.pop()$}
        \State{$mid \gets$ binary\_search$(kvs.keys, begin, end, node.key)$}
        \If {$node.left=null$} 
          \State{create\_sub\_tree$(kvs[begin:mid])$}
          \State\Comment{construct a new sub tree}
          \State{$node.left \gets sub\_tree$}
        \Else
          \State{$queue.push(<begin, mid, node>)$}
        \EndIf
        \If {$node.right=null$} 
          \State{create\_sub\_tree$(kvs[mid:end])$}
          \State{$node.right \gets sub\_tree$}
        \Else
          \State{$queue.push(<mid, end, node>)$}
        \EndIf
      \EndWhile
\EndProcedure
\end{algorithmic}

    \end{algorithm}
  \end{minipage}
\end{figure}

The \texttt{vector\_insert} in Algorithm~\ref{algo:vector} shows  \texttt{vector\_write}, one vector operation, in a binary search tree following the Gather-Apply paradigm.
It firstly reads the information to decide where to insert these nodes and then applies these insert in the correct position.
Without batching, two read rounds are needed if insert operation $A$ and $B$ read the same node. 
When we execute $A$ and $B$ with one \texttt{vector\_write} operation, it only needs one round read to access this node.
Similarly, if several operations modify the same NVM memory, they will be compacted to one NVM write in \texttt{vector\_insert}. 

The caching and batching optimization 
described for tree-like data structure can also be applied to skip-list. Specifically, higher degree nodes in skip-list will be cached. Vector operation (containing sorted operations) for skip-list can similarly reduce the number of \texttt{RDMA\_Read} calls.

\section{NVM Data Management}
\label{sec:mm}

\subsection{Back-end Interface and Metadata}
\label{sec:back_mm}

At the back-end nodes, we implement 
NVM management APIs since using only one-sided RDMA 
operations is inefficient.
In addition, since they provide the basic functions 
needed by all applications,
it is convenient to support them directly 
in the back-end nodes to reduce the network communication to only one round for RPC invocation.
In the \anvm framework, two memory management APIs are provided: \texttt{remote\_nvm\_malloc} and \texttt{remote\_nvm\_free}. The front-end node can use them to allocate and release NVM memory in the back-end nodes. To ensure simplicity, we only implement fixed-size memory management.
Moreover, we use a persistent bitmap to record the usage of NVM, with one bit indicating the allocation status of each block. 
The two design decisions ensure fast recovery.
Since front-end nodes connect to the back-end via one-sided RDMA, we use the RFP~\cite{su2017rfp} to implement the interfaces. Because the front-end puts the requests via \texttt{RDMA\_Write} and gets the responses via \texttt{RDMA\_Read}, the back-end 
is passive and does not need to deal with any network operation. It simplifies the implementation.

The back-end nodes also need to store {\em metadata} for recovery since nothing will be left on the front-end after failure. 
In \anvm frameworks, the metadata are stored in the ``well-known'' locations to all front-end and back-end nodes. This is the \emph{global naming} space for recovery~\cite{arulraj2015let}.
After restarting, both 
front-end or the back-end nodes know the location 
to find the needed information/data before recovery. 
Then, the back-end node maps the virtual memory address to the previous NVM mapped regions. With this mechanism, a pointer to the back-end NVM is still valid after restarting, --- this pointer will be mapped to the previous NVM location. 



The following metadata are stored in the global naming space. 
1) During recovery, the front-end nodes need to know the NVM area address belonging to its data structure, and this NVM area, including the data and log area. It is needed for physical to virtual address translation
for the corresponding front-end node. 
2) The front-end nodes need to know the location of data structures. 
It is achieved by storing the root reference
of data structures, e.g., the address of the root node for a tree.
3) The allocation bitmaps indicate whether a block of NVM is allocated. This information is used to reconstruct the memory usage lists and soon recover the back-end allocator. 
4) Addresses of log areas, LPNs (Log Processing Number, indicating
the next entry in the memory log area) and the OPNs (Operation Processing Number, indicating the last operation log whose memory log is still not persistent) are used to find the logs together with the location of the next logs. They can be used for the back-end node to reproduce logs (memory log) and for the front-end node to recover the data structure operations (operation log). 

\subsection{Front-end Allocator}

The design of the front-end allocator is inspired by the slab allocator~\cite{bonwick1994slab}. The back-end allocator provides slabs to the front-end allocator, and the front-end manages these slabs in finer granularity.  
The slabs in the front-end are organized in full/partial/empty list  
according to how much capacity is consumed in the corresponding page. To support finer granularity allocation, we use a simple \emph{best-fit} mechanism in the front-end.
To improve the NVM utilization, a threshold is defined as the maximum free blocks number, and the front-end
nodes will reclaim free blocks periodically. While reclaiming, The front-end nodes send the request to the back-end nodes to free the reclamation slabs via RPC. 
When allocation size is larger than the size of a slab, the front-end node directly allocates 
memory in the back-end using RPC and back-end interface.
\begin{table}\footnotesize
\centering
\caption{Comparison of Different Allocators. We set the slab size as 128 Bytes and 1024 Bytes separately.}
\label{tab:alloc}
\renewcommand{\arraystretch}{0.5}
\begin{tabular}{lll}
\toprule
Type/Tput(MOPS) & Alloc  & Free \\ \midrule

Glibc       & 21.0 & 57.0 \\ \midrule
Pmem        & 1.42 & 1.38 \\ \midrule
RPC allocator         & 0.33 & 0.88 \\ \midrule
\textbf{Two-tier allocator}  (128 Bytes)    & 1.33 & 2.41 \\ \midrule 
\textbf{Two-tier allocator} (1024 Bytes)  & 6.42 & 13.90\\ \bottomrule

\end{tabular}
\end{table}

\textbf{Benchmark.} 
We compare the two-tier allocator of
\anvm framework with persistent allocator and standard Linux \emph{Glibc} allocator.
As table~\ref{tab:alloc} shows, \emph{Glibc} achieves the highest throughput (21.0 MOPS and 57.0 MOPS) but without persistent guarantee.
\emph{Pmem} allocator is a persistent allocator from NVML project~\cite{pmem}, and can reach 1.42 MOPS.
With only the back-end (RPC) allocator,
the throughput is only 23\% and 64\% of Pmem allocator because of the network overhead.
With two-tier allocator, 
the throughput is similar or even better performance than Pmem allocator.

\section{Concurrency Control}\label{sec:concurrency}

This section describes the concurrency control
mechanisms to support SWMR (Single Writer Multiple Reader) access patterns.
Based on applications, 
\anvm can support both lock-free and lock based data structures.

\begin{figure}[!t]
  \centerline{\includegraphics[width=.9\columnwidth]{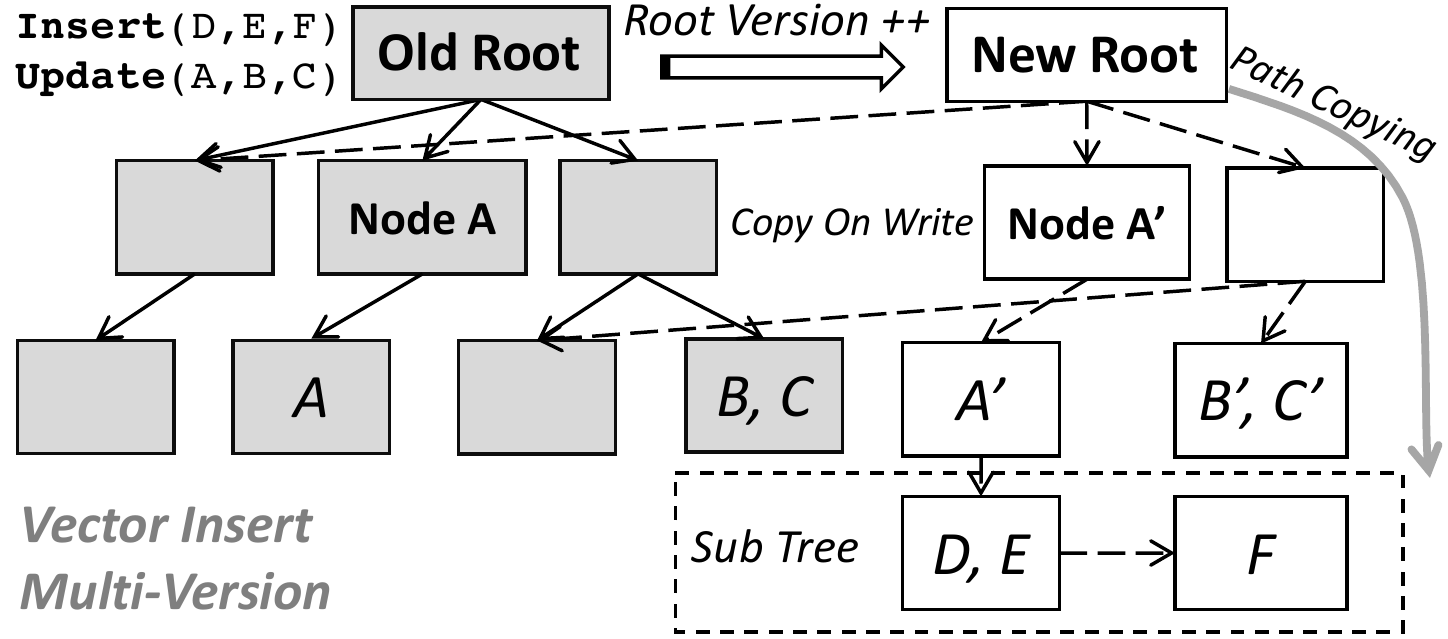}}
\caption{Overall Multi-version Data Structure}
\label{fig:append}
\end{figure}

\begin{figure}[!t]
  \centerline{\includegraphics[width=.85\columnwidth]{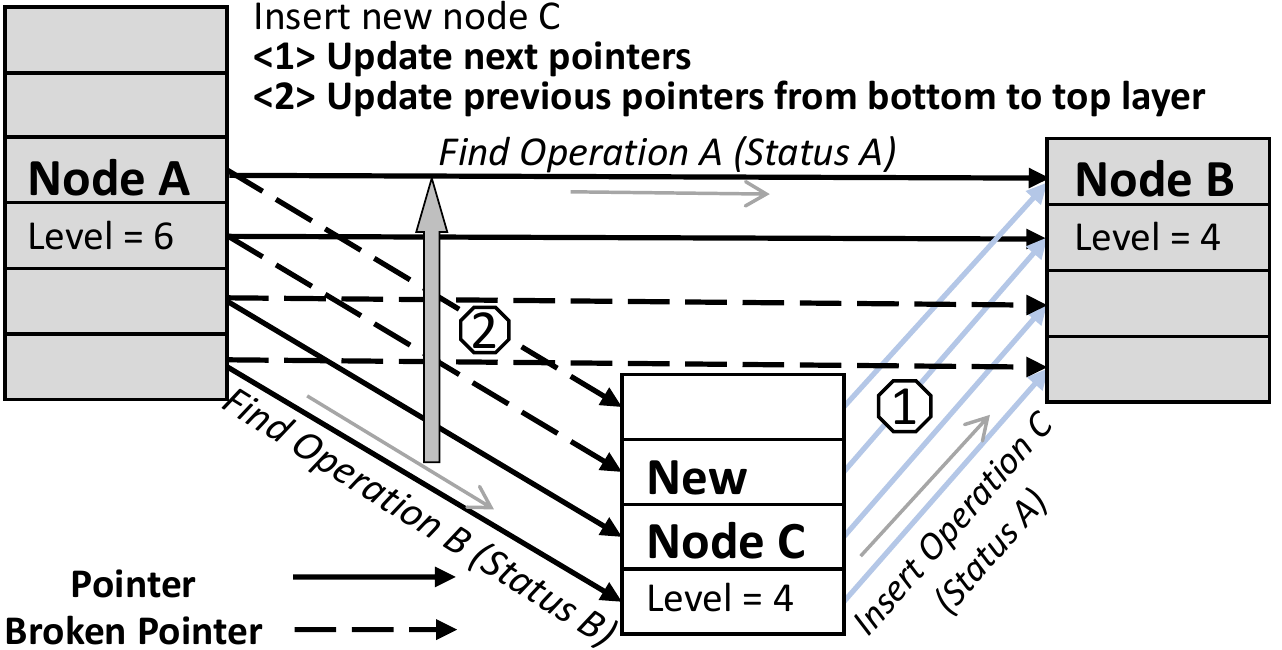}}
\caption{Naturally Lock-Free in Skip-list.}
\label{fig:skiplist}
\end{figure}

\subsection{Exclusive Write}

Under the SWMR mode, write operations are exclusive. 
Therefore, the writer should acquire an exclusive lock first. If it succeeds, it fetches the LPN 
(Log Processing Number, refer to Section~\ref{sec:back_mm} for
its management), and then executes the write operation. After finishing appending logs to the remote NVM based on LPN, it should release this exclusive lock. While the exclusive writes 
are being performed, other write operations (if any) will be blocked until the current writer has completed the current write operation.

\algdef{SE}[DOWHILE]{Do}{doWhile}{\algorithmicdo}[1]{\algorithmicwhile\ #1}%

\begin{figure}[!t]
  \centering
  \begin{minipage}{1.0\linewidth}
    \begin{algorithm}[H]\footnotesize
\caption{Writer Lock}\label{algo:write}
\begin{algorithmic}[1]

\Procedure{Writer Lock}{}
	\State{\textbf{while} $rdma\_compare\_and\_swap(\textbf{L}, Locked) = Locked$}
\EndProcedure

\Procedure{Writer Unlock}{}
	\State $rdma\_atomic\_write(\textbf{L}, UnLocked) $
\EndProcedure

\end{algorithmic}
\end{algorithm}
  \end{minipage}
\end{figure}

As shown in Algorithm~\ref{algo:write}, for
\texttt{Write\_Lock}, we 
leverage the RDMA atomic verbs, \texttt{RDMA\_Compare\_And\_Swap}~\cite{yoon2018distributed} 
to implement it as a distributed spin-lock. When releasing the lock, the writer resets it via a \texttt{RDMA\_Read}. In \anvm framework, 
to handle the failures while holding the lock, 
every write lock acquire/release operation should write a record (lock-ahead log) to the back-end node before appending the memory logs. Thus, if the front-end crashed before releasing the lock, we can identify the lock need to be released during recovery.

\subsection{Lock-Free Data Structure}

\textbf{Multi-version Data Structure.} 
Our design of lock-free tree-like data structure
is inspired by append-only B-tree~\cite{anderson2010couchdb, appendonlybtree} and persistent data structures~\cite{driscoll1989making, okasaki1999purely}. 

Multi-version is a widely used method in optimistic concurrent control~\cite{becker1996asymptotically, sowell2012minuet}.
Multi-version data structures will first make copies of the corresponding data if needed. Then the data will be modified or new data items are inserted. For example in Figure \ref{fig:append}, the writer copies all the affected nodes along the path to the root, a.k.a., path copying~\cite{rodeh2008b}. Then, the nodes in the path will update some of the pointers 
pointing to the old data. Finally, the data will be inserted into the new path. After finishing all these operations, the root will be atomically 
changed to the new root by updating the root pointer. \textit{Vector operation} discussed in Section~\ref{sec:ds_opt} can help here to reduce the number of network round trips significantly. Since the readers can always get consistent data, this kind of concurrent control does not affect 
the performance of readers. 

\textbf{Skip-List: Naturally Lock-Free.}
Some data structures like skip-list are naturally lock-free and the only concern is to carefully choose the order of operations~\cite{fomitchev2004lock, crain2013no}. As shown in Figure~\ref{fig:skiplist}, the writer first creates the newly allocated node and sets the pointers in the new node accordingly (\textcircled{1}). After that, the previous pointers will be updated from the bottom to the top (\textcircled{2}). Readers can still get 
(potentially different) consistent views of Skip-list in such scenario, thus, the lock is not 
required~\cite{herlihy2006provably}.

\textbf{Lightweight Recovery. }
In the multi-version data structure, the only in-place update is the root pointer. 
However, the pointer changing is atomic. Therefore, it doesn't need a recovery process as the discussion in ~\cite{arulraj2015let}. While recovering, the front-end can use the root pointer (which is well-known via naming mechanism) to find out the whole data structure.

\textbf{NVM reclamation. }
The use of lock-free data structures needs 
to ensure that memory is safely reclaimed, which further complicates the garbage collection\cite{fraser2004practical}. In 
\anvm framework, this requirement can be achieved by a lazy garbage collection mechanism. After version changes, the front-end should release the old version's data. Back-end delays this operation for $n$ $\mu s$ and then reclaims corresponding memory. It requires that the latency of each 
pending data structure operation should be less than $n$ $\mu s$ to avoid memory leak (i.e., access the reclaimed memory). 
\subsection{Lock Based Data Structure}

\textbf{Write Preferred Lock.}
RDMA library provides atomic verbs which is an appropriate way to implement distributed sequencer~\cite{kaminsky2016design, yoon2018distributed}.
Algorithm~\ref{algo:lock} shows 
the implementation of 
retry-based read locks by using the sequence number (SN), an 8 bytes integer variable. 
Distinct from Algorithm~\ref{algo:write}, which is 
invoked by the front-end nodes, 
\texttt{Write\_Begin} and \texttt{Write\_End} is executed by the back-end nodes.
When a back-end node applies the 
persisted memory log to the real 
data structure in NVM, it atomically
increases the SN twice before and 
after the modification.
\texttt{Reader Lock} and \texttt{Reader Unlock}
are invoked by front-end nodes before
and after a sequence of reads.
To disallow reads when data are being updated, 
it needs to wait until the current SN is odd.
To ensure reads in between 
get the consistent view,
\texttt{Reader Unlock} needs to check that 
SN is unchanged since \texttt{Reader Lock}.
If the data are inconsistent, the readers need to \emph{retry and fetch} the data again.

\textbf{Lock Benchmark.} 
We make a ping-point test about the lock's performance as in Frangipani~\cite{thekkath1997frangipani}. In our test scenario, six readers and one writer try to access the same data in the back-end and the writer's workload is 10 \% write and 90 \% read. The results show that each reader's average throughput is 260 KOPS (1.56 MOPS in total) and writer's throughput is 539 KOPS, separately.
The reader's failed ratio (i.e., a try for reading data is failed) is only 3 \%. When setting the workload as 50 \% write, reader's throughput will 
drop to only 165 KOPS with a 26 \% fail ratio. The write-preferred lock makes writers to gain a higher throughput than readers.



\algdef{SE}[DOWHILE]{Do}{doWhile}{\algorithmicdo}[1]{\algorithmicwhile\ #1}%



\begin{figure}[!t]
  \centering
  \begin{minipage}{.9\linewidth}
    \begin{algorithm}[H]\footnotesize
\caption{Writer Preferred Lock}\label{algo:lock}
\begin{algorithmic}[1]

\Procedure{Write Begin}{}
	\State $gcc\_atomic\_increment(\textbf{SN})$
\EndProcedure

\Procedure{Write End}{}
	\State $gcc\_atomic\_increment(\textbf{SN})$
\EndProcedure

\Procedure{Reader Lock}{}
	\Do \State $\textit{ret}  \gets rdma\_atomic\_read(\textbf{SN})$
	\doWhile{$\textit{ret} \ is \ odd$}
	\State $\textit{start\_sn} \gets ret $
\EndProcedure

\Procedure{Reader Unlock}{}
	\State \Return $\textit{start\_sn} \neq  rdma\_atomic\_read(\textbf{SN})$
\EndProcedure

\end{algorithmic}
\end{algorithm}
  \end{minipage}
\end{figure}

\begin{table*}\footnotesize 
\centering
\caption{Performance Improvement and Comparison (in KOPS) (R: using log reproducing, C: caching 10\% NVM size in the front-end, B: batching with batch size 1024. The evaluation uses the one-to-one setting with 100\% write workloads harnessing all optimization.
Reasons for \emph{empty cells}: Data structure with time complexity $O(1)$ (i.e., HashTable/SmallBank) cannot apply batching optimization.
In Queue/Stack implementation, batch and cache should be combined together.
}


\label{tab:opt}
\renewcommand{\arraystretch}{-1000}
\begin{tabular}{ lllllllllll }
\toprule
	 & TX(SmallBank) & TX(TATP) & Queue & Stack & HashTable & SkipList & BST & BPT & MV-BST & MV-BPT \\ \midrule
    Symmetric & 654 & 214 & 1199 & 1087 & 1097 & 125.2 & 84.5 & 305.2 & 42.2 & 18.6 \\ \midrule
	Symmetric-B & - & 260 & 2279 & 2255 & - & 209.0 & 151.0 & 343.0 & 146.1 & 76.0 \\ \midrule
	\anvm-Na\"ive & 254 & 10.2 & 301 & 285 & 315 & 5.0 & 19.0 & 11.5 & 7.0 & 7.4 \\ \midrule
	\anvm-R & 295 & 12.4 & 833 & 828 & 385 & 7.7 & 22.9 & 13.7 & 12.3 & 9.8 \\ \midrule
    \anvm-RC & 362 & 63.7 & - & - & 445 & 40.4 & 59.5 & 77.1 & 28.4 & 17.8 \\ \midrule
	{\bf \anvm-RCB} & - & 127.5 & 1678 & 1449 & - & 66.0 & 134.2 & 184.3 & 88.9 & 60.2 \\ \bottomrule

\end{tabular}
\end{table*}



\begin{figure}[!t]
  \centerline{\includegraphics[width=.9\columnwidth]{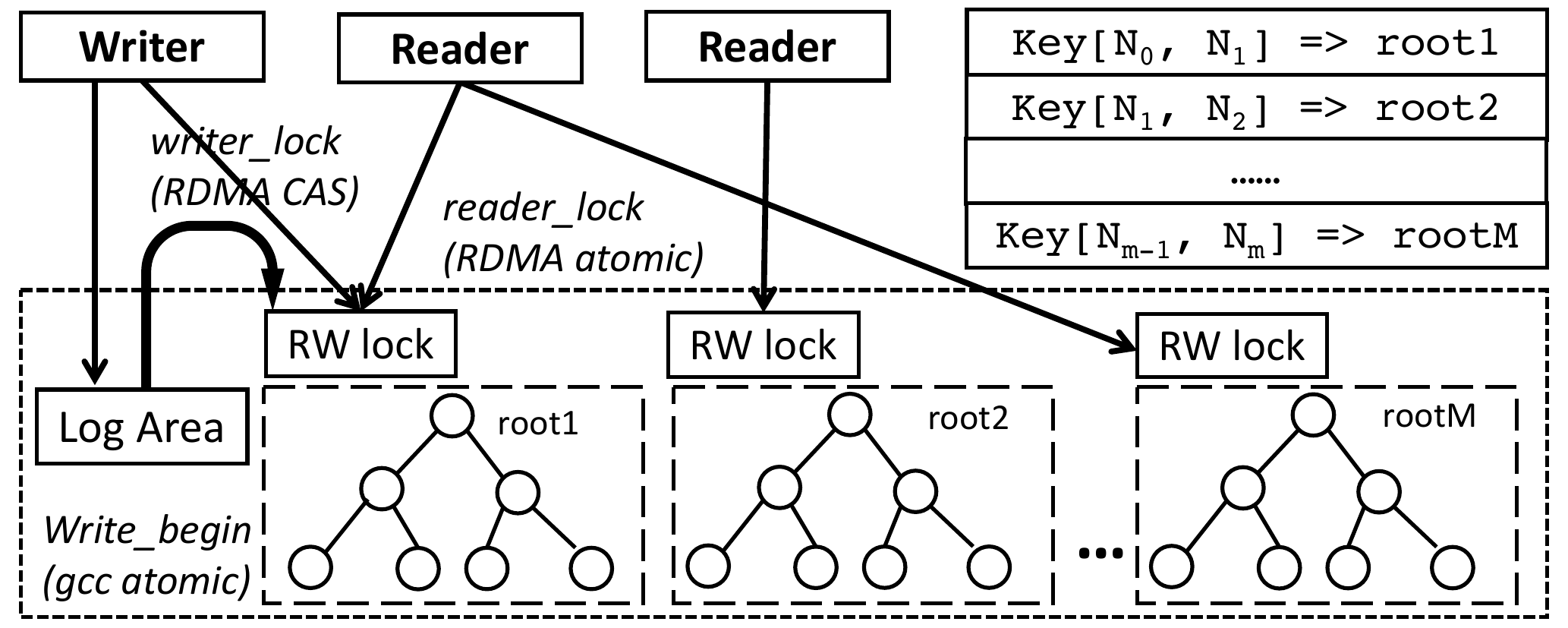}}
\caption{Data Structure Partition.}
\label{fig:partition}
\end{figure}
\subsection{Data Structure Partition}

We use partitioning to eliminate the potential
bottleneck due to the lock, to achieve both high throughput and better scalability~\cite{tu2013speedy, lim2014mica}.
Similar to the support of large size data structure in Section~\ref{sec:batching}, 
\anvm framework adopts key-hashing partitioning to improve the performance of various data structures.
As shown in Figure~\ref{fig:partition}, Each partition has its own write lock and index data structure.
While the writer is executing write operation in one of the partitions, multiple readers can still concurrent access other partitions.
In our implementations, we always separate the data structure into four partitions to simplify the evaluation.

 Lock-free data structures benefit the reader but create multiple copies by writers. Lock-based data structures 
prioritize the writer without extra copies, but readers have to read multiple times until consistent data are obtained. The right choice depends on specific applications. 



\section{Recovery and Replication}
\label{sec:recovery}
\subsection{Replication}\label{sub:rep}

Recent work Mojim (based on symmetric architecture)~\cite{zhang2015mojim} adopts a primary-mirror architecture to avoid complex protocols for achieving availability. The \anvm framework similarly needs {\em at least} one mirror-node attached with the
non-volatile device like SSD, Disk or even NVM. 
To improve fault tolerance, it is preferred to deploy 
mirror-nodes to different racks.  
The back-end nodes replicate the memory/operation logs to mirror-nodes before committing the transaction and
acknowledging the front-end node to avoid high overhead. 
If the mirror node is equipped with NVM, the mirror node also implements 
a log replay function similar to the back-end to 
apply logs to the replicated real data structure.
Replicated logs in mirror nodes are read-only. 
When the back-end node crashes, 
if the mirror node is equipped with NVM, 
it will be voted as the new back-end. 
Otherwise, the front-end nodes 
use the logs and data structures from the mirror node 
to recover the data structure to a new back-end node
with the NVM device.

In our implementation, the back-end node is responsible for ensuring that the replica is persistent in its mirror-node. The front-end node only needs to ensure that 
data is stored in the back-end NVM, but does not 
wait for an acknowledge after replication completes. 
Thus, the replication phase is performed asynchronously.

\subsection{Data Structure Recovery}\label{subsub:recover}


With the log mechanism, \anvm framework
ensures crash consistency with the non-volatile data 
and logs stored in the back-end node. 
This section discusses the details of 
different recovery scenarios based on the 
failure of different components. 

Similar to most distributed systems, 
we implement a consensus-based voting system, i.e., etcd~\cite{etcd} or ZooKeeper~\cite{hunt2010zookeeper}, to detect machine failures. Leases are used to identify whether a node, i.e., front-end or back-end, is still alive or not. If the lease expires and the node cannot renew its lease, the node is considered to be crashed. 
We implement this mechanism as {\em keepAlive service}.

{\bf Case 1: Front-end reader crash. } 
If the front-end node crashes when performing 
a read, it only needs to gain the meta-data via naming mechanism and resume execution after rebooting.

{\bf Case 2: Front-end writer crash.}
If the front-end node crashes when performing a write, 
the back-end will know this information through 
keepAlive service. 
After the front-end node reboots from crashing, 
if there still exists memory logs not replayed from the
front-end node, the back-end node 
will validate whether all log entries of 
the last transaction are flushed to the NVM
or not via checksum.
If this transaction log is consistent {\bf (Case 2.a)}, 
the back-end will notify the front-end to resume as normal,
same as the case of reader crashing. 
Otherwise {\bf (Case 2.b)}, the back-end will notify the front-end
that the last transaction log is inconsistent. Thus, 
the front-end node will fetch the LPN, OPN and operation logs, 
of which memory logs is not replayed, 
and then re-executes the uncommitted transaction according to the operation log.
{\bf (Case 2.c)} In most cases, there are several operation logs whose corresponding memory logs are not flushed to back-end yet. the front-end will process as {\bf (Case 2.b)}. 


{\bf Case 3: Back-end transient failure. }
When the back-end node fails while executing \texttt{RDMA\_write/read}, the front-end can detect it 
through the feedback from RNIC, i.e., the destination is un-reached.
Then, it will wait for the notification for
the back-end node recovery or a new voted back-end. 
After rebooting, the back-end node will first reconstruct the mapping between the physical addresses and virtual addresses. 
The reconstruction is possible because such mapping is also stored in NVM and its beginning address is well-known via \emph{global addressing} schema as the description in Section~\ref{sec:back_mm}. 
After that, the back-end node 
checks whether the last transaction log is consistent.
If there is no transaction/operation logs left, 
or 2) the transaction log is consistent {\bf (Case 3.a)},
the back-end can start its normal execution immediately, 
i.e., reproducing memory logs to data structures if any log has not been applied, and then notify its liveness to the front-end
nodes. 
If the transaction log is inconsistent {\bf (Case 3.b)}, 
the back-end node will notify the corresponding front-end
nodes about its crash, and the front-end will flush the memory logs again to redo this transaction.
It is possible since the front-end node must have not 
received the persistent acknowledgement.
If existing operation logs are ahead of current 
memory logs {\bf (Case 3.c)}, which means that
the memory logs of several operation logs have not 
been flushed from front-end due to batching,
the back-end node will notify the front-end about this, 
and the front-end will continue to execute the next operation.

{\bf Case 4: Back-end permanent failure. }
In this case, one of the mirror nodes will be voted as the new back-end and provides service to the front-end. 
The new back-end node will broadcast to living front-ends to announce such information. 
After that, the front-end will reconstruct the data structures to a new back-end by using the data and logs in the mirror nodes.

{\bf Case 5: Mirror node crash. }
The consensus-based based service will detect the failure and remove it out of the group.

If both front-end and back-end crash, the keepAlive service will coordinate front-end and back-end nodes, and let the back-end nodes to recover first. They will first check the status as
in {\bf Case 2}. After that, the front-end will determine how to recover according to the back-end's failure cases
in {\bf Case 1}.

\section{Evaluation}\label{sec:eva}

Our evaluations attempt to answer the following questions:

\textbf{\RNum{1}.} How does \anvm perform, --- how it is compared to symmetric setting and naive asymmetric implementation? \textbf{\RNum{2}.} How much performance improvement can 
batching and caching deliver?
\textbf{\RNum{3}.} How does \anvm perform under multiple front-end nodes?
\textbf{\RNum{4}.} How does \anvm perform under different workloads?

\subsection{Evaluation Setup}\label{sub:imp:emu}
\noindent\textbf{Hardware Setup.} The experiment cluster contains ten machines, each of which is equipped with an 8-cores CPU (Intel Xeon E5-2640 v2, 2.0 GHz), 
96 GB memory, and one Mellanox ConnectX-3 InfiniBand with network bandwidth of 40Gbps. 
Up to three machines are used 
as the back-end nodes or mirror nodes.


\noindent\textbf{NVM Emulator.}
We use 60 GB DRAM as remote NVM device, and 6GB DRAM as the front-end DRAM for caching data. 
Similar to prior works~\cite{bhandari2016makalu, zhang2015mojim, liu2017dudetm}, 
we set the write latency as 200$ns$ and read latency as the latency of DRAM. This is due to the read/write asymmetry in NVM.

\begin{figure}[h]
\renewcommand{\arraystretch}{0.5}
\begin{tabular}{cc}
\subfloat[Lock-Free Data Structure]{\includegraphics[width=0.5\columnwidth]{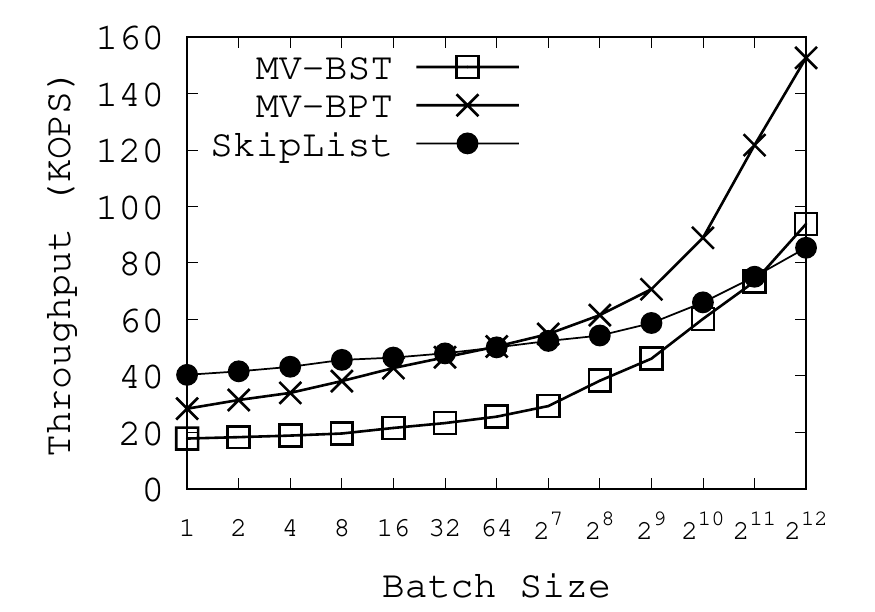}}
\subfloat[Lock Based Data Structure]{\includegraphics[width=0.5\columnwidth]{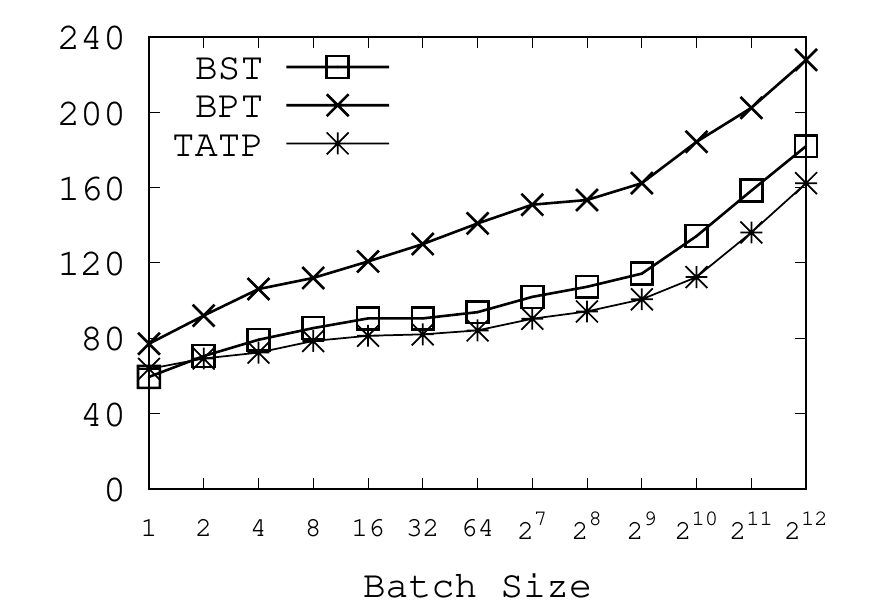}} \\
\end{tabular}
\caption{Throughput with different batch sizes}\label{fig:batch-size}
\end{figure}

\begin{figure}[h]
\begin{tabular}{cc}
  \subfloat{\includegraphics[width=0.66\columnwidth]{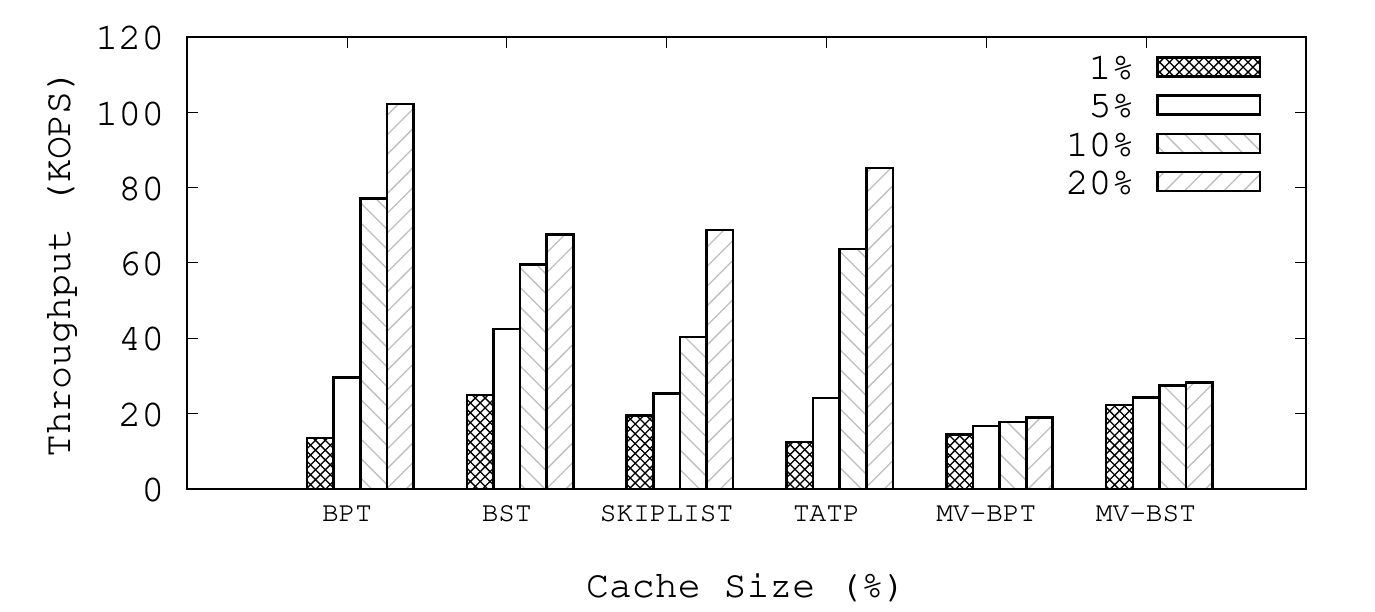}}
  \subfloat{\includegraphics[width=0.348\columnwidth]{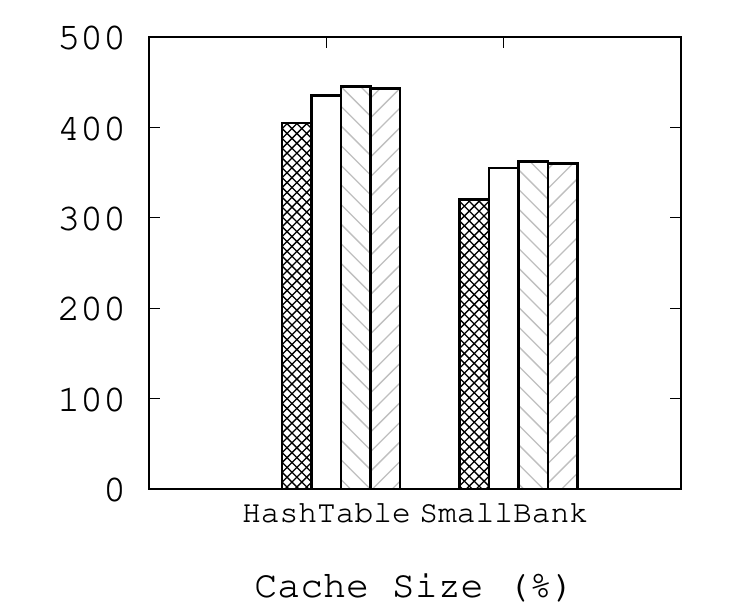}}
\end{tabular}
\caption{Throughput with different cache sizes}
\label{fig:cache-size}
\end{figure}

\begin{figure}[h]
\renewcommand{\arraystretch}{0.5}
\begin{tabular}{cc}
\subfloat[Lock-Free Data Structure]{\includegraphics[width=0.5\columnwidth]{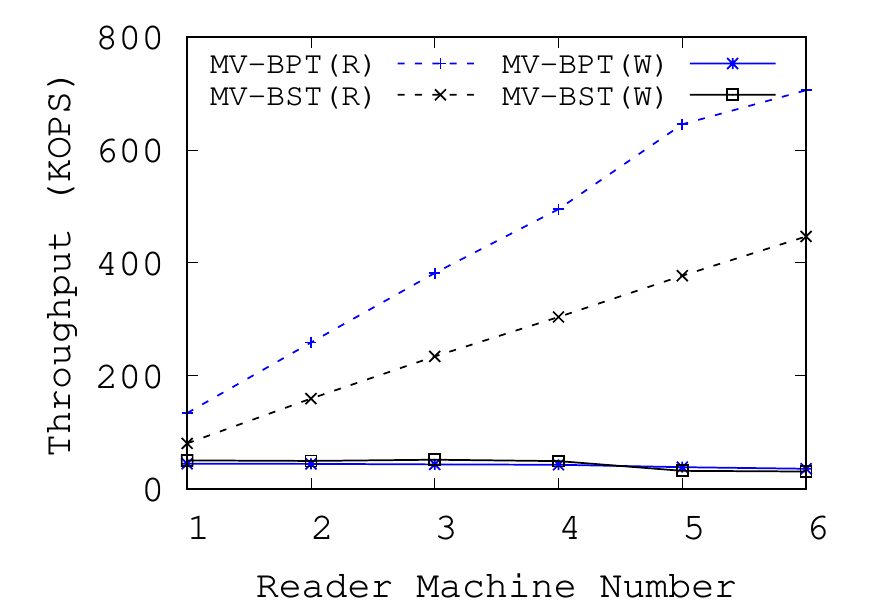}\label{fig:scale-up-a}}
\subfloat[Lock Based Data Structure]{\includegraphics[width=0.5\columnwidth]{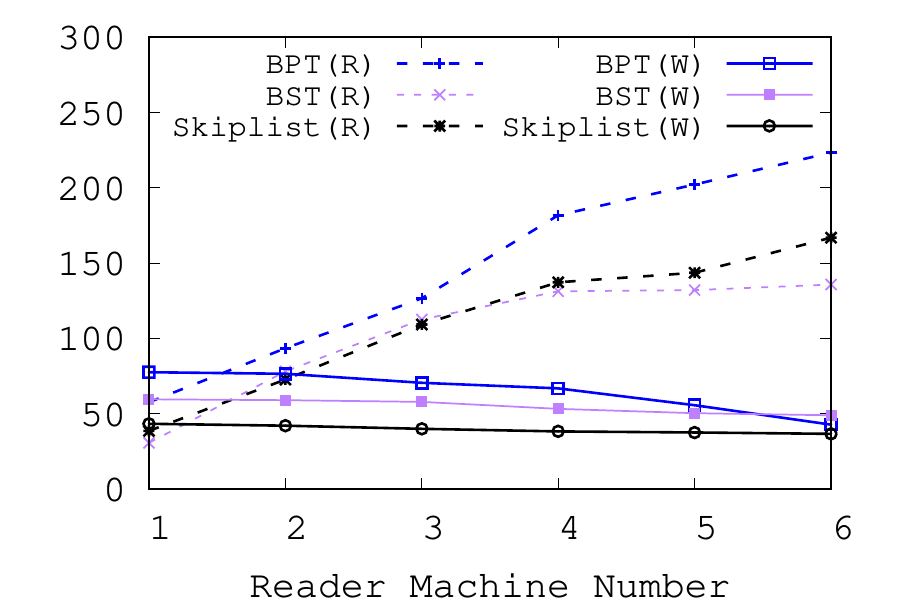}\label{fig:scale-up-b}} \\
\end{tabular}
\caption{Scalability of multiple readers. The workload of the writer is 100\% insert. 
R/W represents reader/writer.}\label{fig:scale-up}
\end{figure}

\begin{figure}[h]
  \centerline{\includegraphics[width=0.8\columnwidth]{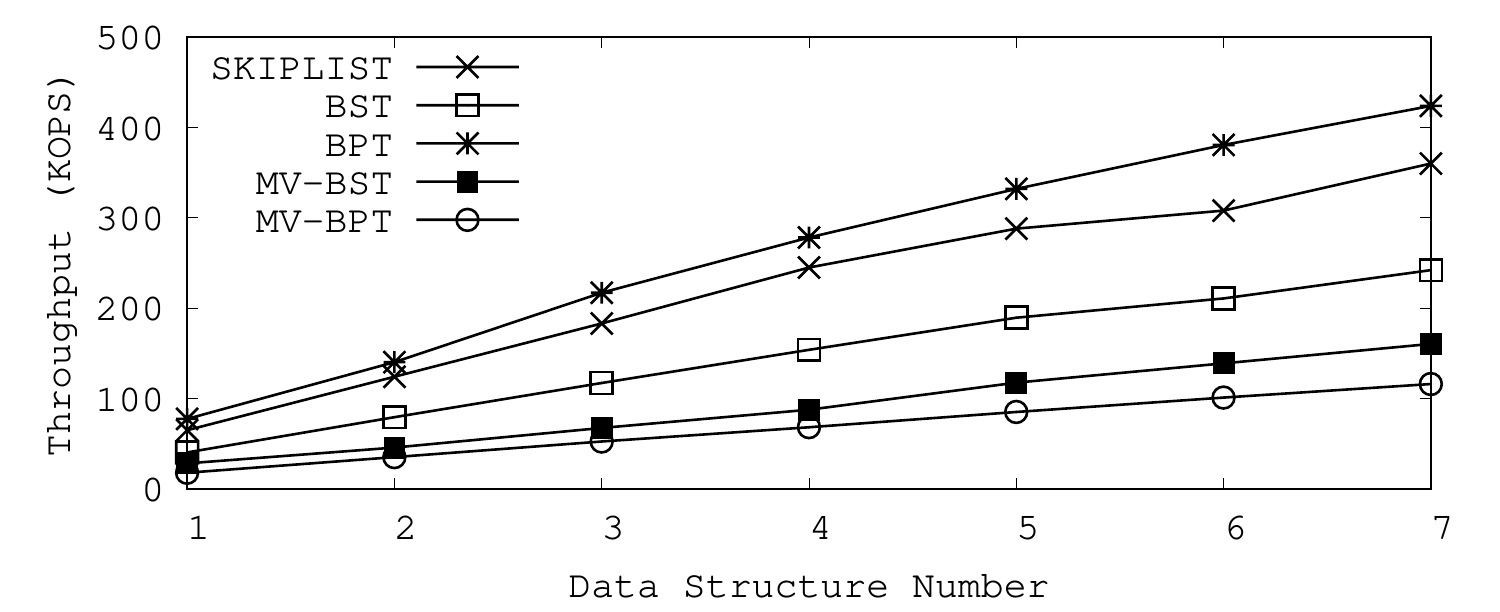}}
\caption{Throughput of multiple data structures in the Same Back-end Machine }
\label{fig:scale-out}
\end{figure}

\begin{figure}[h]
  \centerline{\includegraphics[width=1.0\columnwidth]{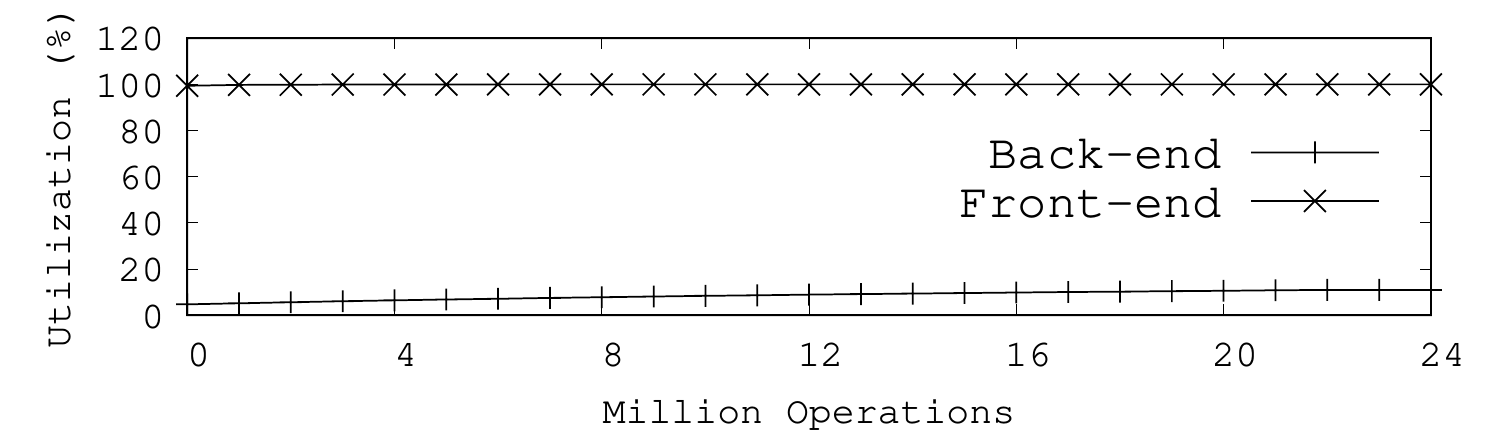}}
\caption{CPU utilization with the operation increasing in BST.  The workload is 10\% put and 90\% get.}
\label{fig:cpu}
\end{figure}

\begin{figure*}[h]
\renewcommand{\arraystretch}{0.5}
\begin{tabular}{cccc}
   \subfloat[BST]{\includegraphics[width=0.48\columnwidth]{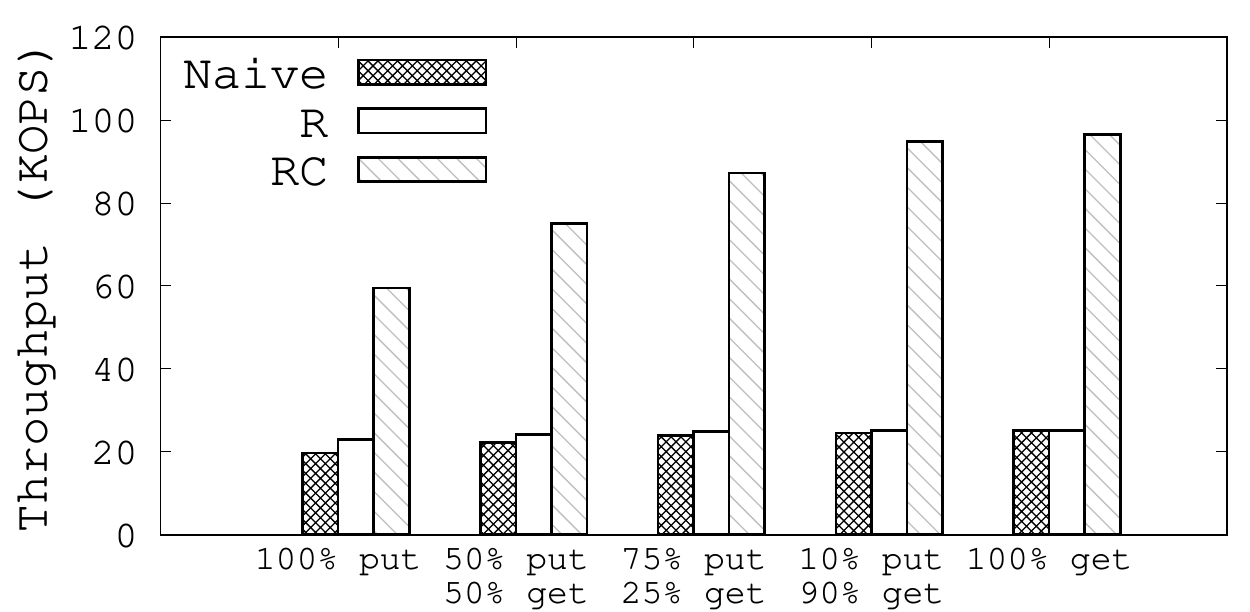}}\hfill
   & \subfloat[MV-BST]{\includegraphics[width=0.48\columnwidth]{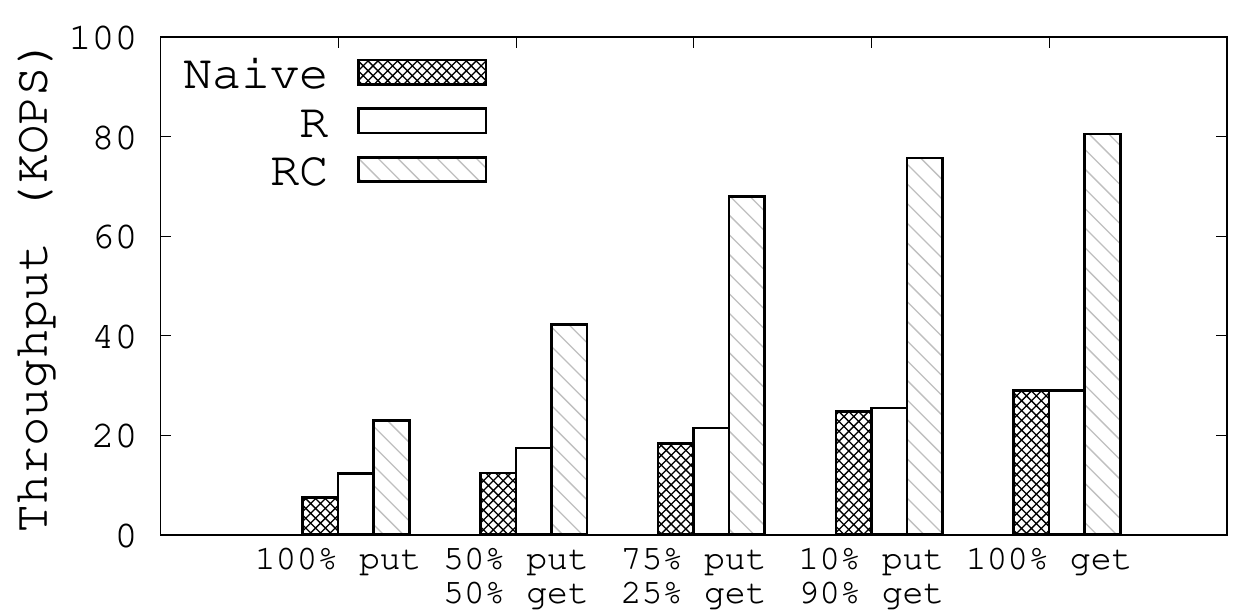}}\hfill
   & \subfloat[BPT]{\includegraphics[width=0.48\columnwidth]{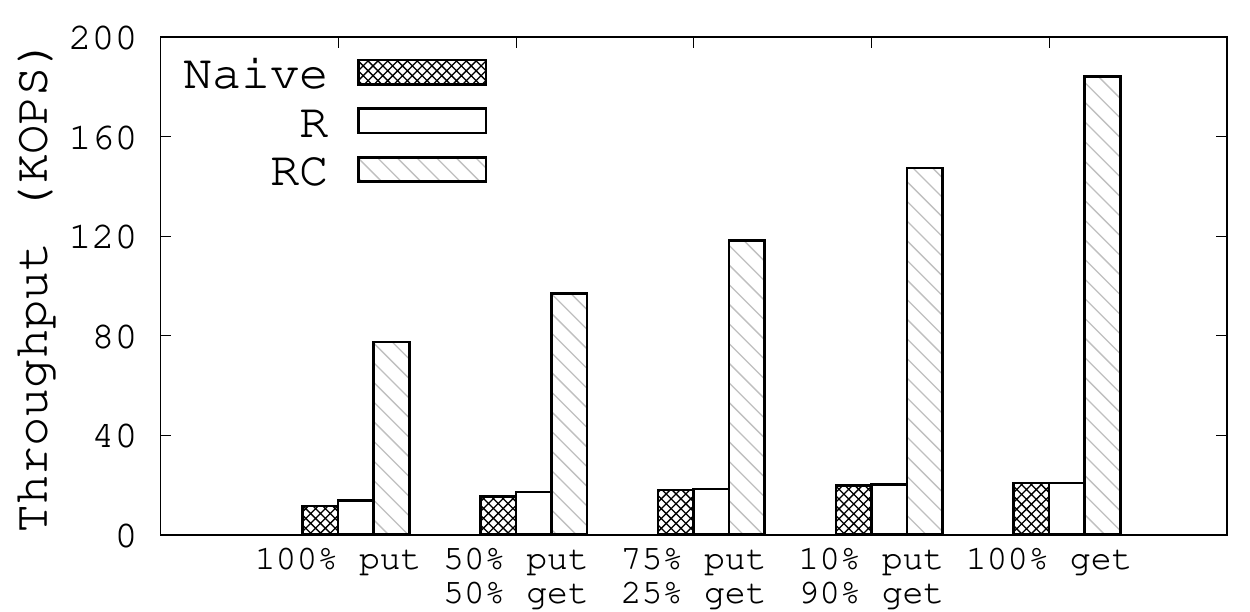}}\hfill
   & \subfloat[MV-BPT]{\includegraphics[width=0.48\columnwidth]{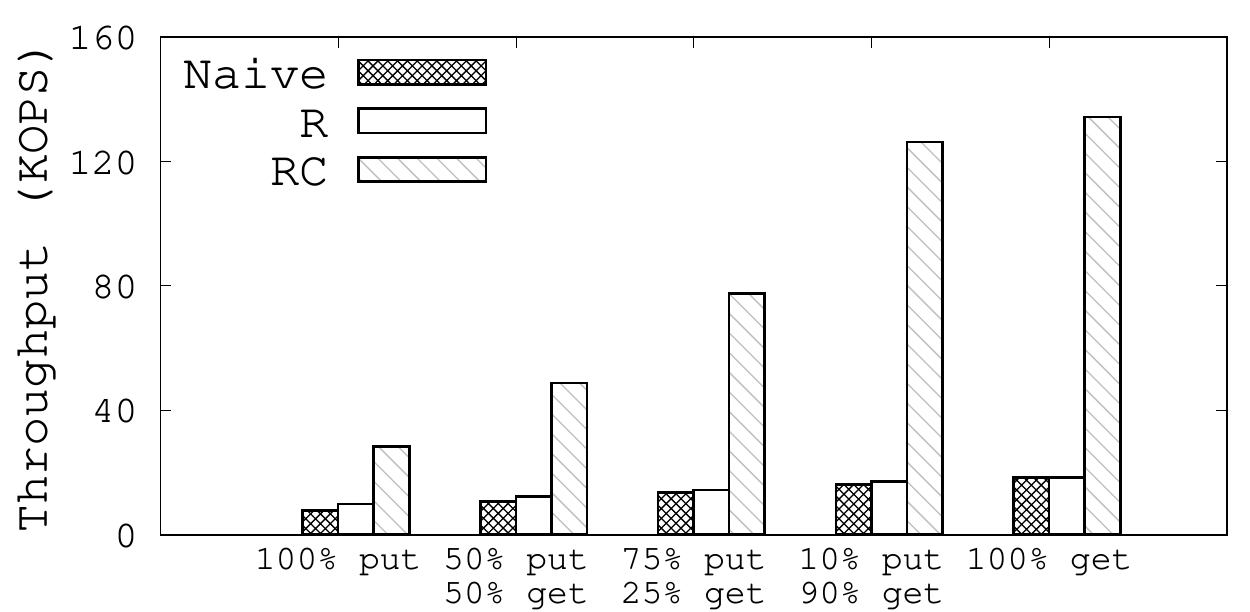}} \\
\end{tabular}
\renewcommand{\arraystretch}{0.5}
\begin{tabular}{cccc}
   \subfloat[SkipList]{\includegraphics[width=0.48\columnwidth]{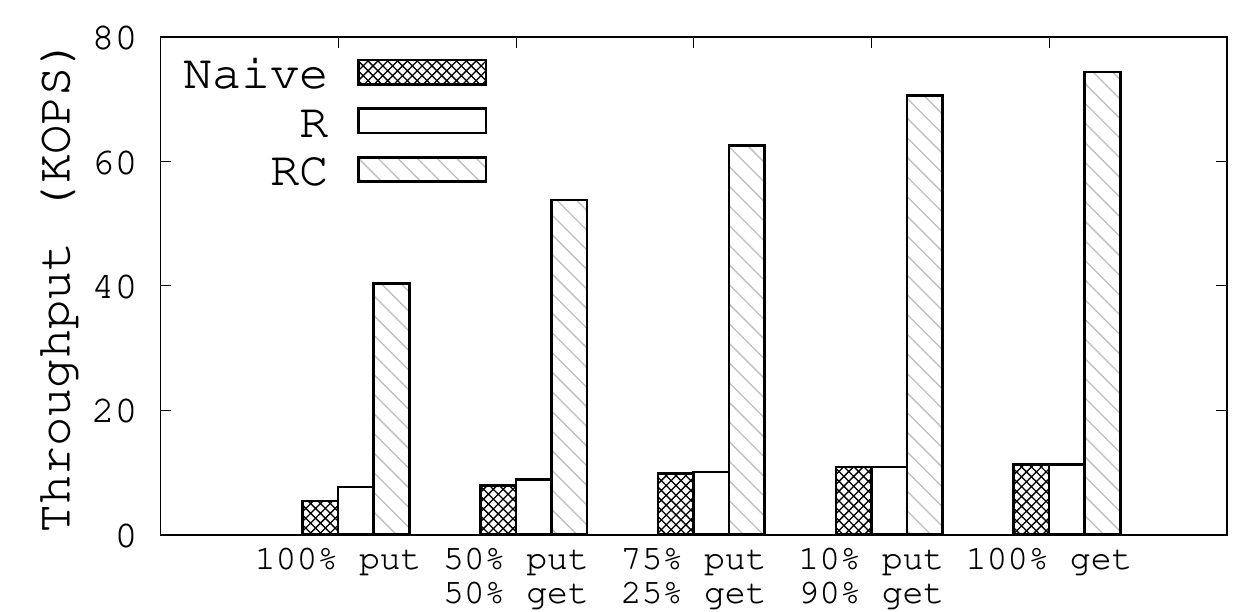}}\hfill
   & \subfloat[Queue]{\includegraphics[width=0.48\columnwidth]{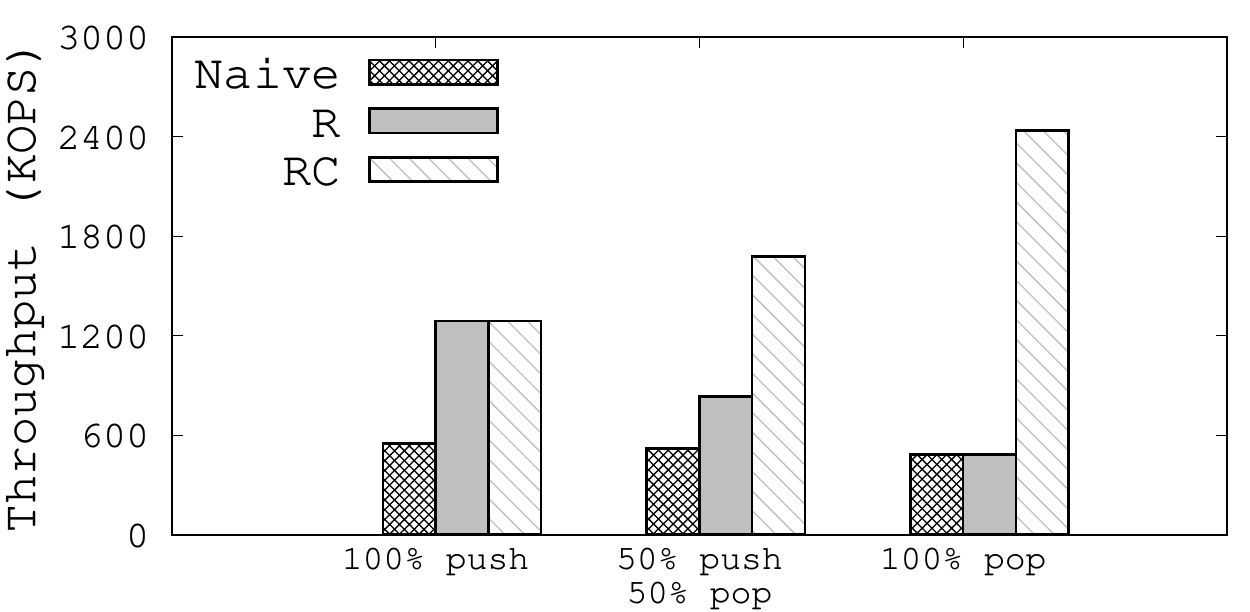}}\hfill
   & \subfloat[Stack]{\includegraphics[width=0.48\columnwidth]{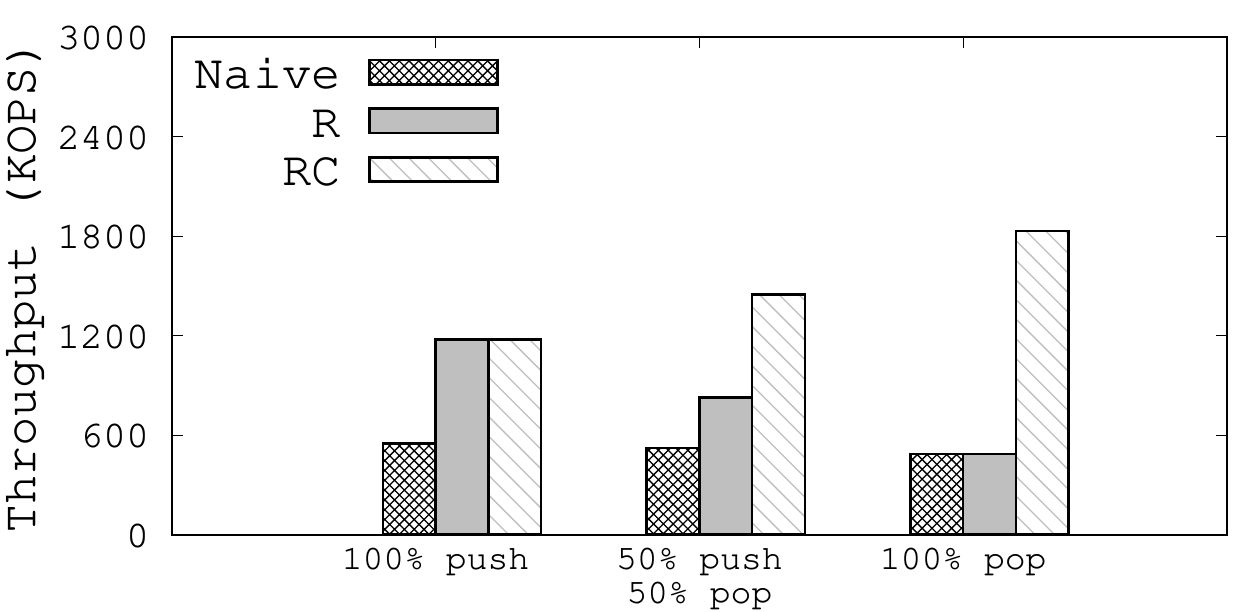}}\hfill
   & \subfloat[HashTable]{\includegraphics[width=0.48\columnwidth]{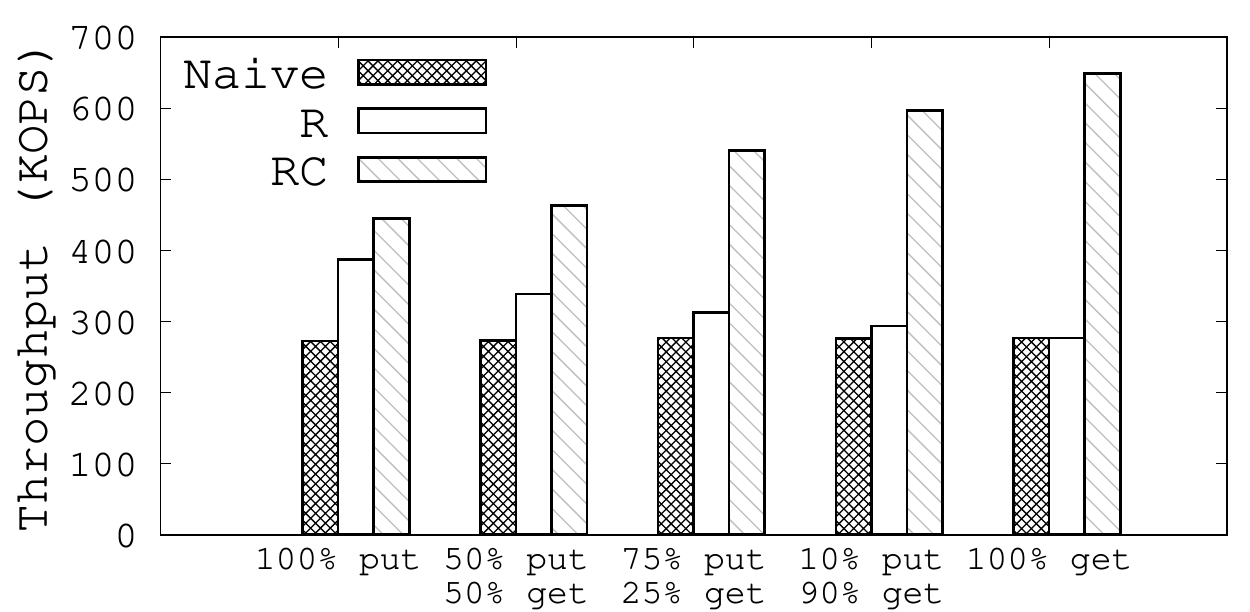}}\\
\end{tabular}

\caption{Throughput with Different Workloads (100\%put, 50\%put+50\%get, 25\%put+75\%get, 10\%put+90\%get, 100\%get)}\label{fig:workload}
\end{figure*}




\subsection{\anvm Performance}
We implement eight widely-used data structures covering 
different access time complexity ($O(1)$ and $O(log (n))$): 
stack, queue, hash-table, skip-list, binary search tree (BST), B+tree (BPT), multi-version binary search tree (MV-BST), and multi-version b+tree (MV-BPT). 
To simplify the evaluations, the key and value are all 64 bits. 
In addition, we use two applications: TATP and SmallBank.


Table~\ref{tab:opt} shows the overall performance as well as the comparison to symmetric and naive implementations. 

\textbf{Compare to Na\"ive Implementation.} 
The na\"ive implementation access remote NVM directly
using RDMA reads and writes without any optimizations. 
The complete implementation denoted as \anvm-RCB (with log \textbf{R}eproducing, \textbf{C}atching, and \textbf{B}atching) can provide nearly  6$\sim$22 $\times$ improvements
compared to na\"ive implementation. 
We see that the cachingis more effective
than other optimizations 
(nearly 2$\sim$7 $\times$ performance improvement).
The reason is that the \texttt{RDMA\_Read} operation
causes the major overhead in accessing a data structure, 
and catching largely eliminates this read overhead.

\textbf{Compare to the Symmetric Setting.} 
We implement the symmetric NVM architecture
by storing data structures in local NVM 
and storing logs in remote NVM for fault tolerance. 
The logs are flushed asynchronously (without waiting for the acknowledgement from remote nodes).
It reaches the upper-bound performance of symmetric NVM architecture, but will obviously cause inconsistency. 
From the results, we see that, \anvm-RCB 
still achieves comparable performance 
to the optimistic performance of symmetric NVM data structures
without consistency. 
Especially, in a few cases (i.e., Queue, Stack, BST, MV-BST, MV-BPT), the performance of \anvm-RCB 
is even better than symmetric NVM without batching.
This is mainly due to the 
small DRAM cache in the front-end nodes. 


\textbf{End-to-end Performance.}
We evaluate application performance by two 
transaction benchmarks: SmallBank~\cite{smallbank} and TATP~\cite{tatp}. We use HashTable and BPT as the index data structure of SmallBank and TATP, separately. As shown in Table~\ref{tab:opt}, the results show \anvm can improve the throughput to 1.42$\times$ in SmallBank and 12.5$\times$ in TATP.

\noindent\textbf{Cost Comparison.}
In the symmetric setting with $m$ machines, it needs $n_1=max(\sum_{i=1}^{i=m}\left\lceil \nicefrac{S_i}{S_0} \right\rceil, m)$ NVM devices (assuming each NVM capacity is $S_0$, the real usage of each NVM is $S_i$). Besides, \anvm needs $n_2=\left\lceil \sum_{i=1}^{i=m}S_i\right\rceil$ NVM devices and
$n_2 \ll n_1$. As we mentioned in Section~\ref{sec:intro}, each server only needs smaller capacity less than $S_0$, thereby the necessary NVM $n_2$ will be fewer than $n_1$ ($n_1 = m$).

\subsection{CPU Utilization}\label{sub:cpu-uti}

Figure~\ref{fig:cpu} shows CPU utilization of front-end and back-end nodes. 
The front-end node keeps running with nearly 100\% CPU utilization but the request only incurs very small CPU usage (4\%$\sim$10\% CPU utilization). 
It matches the intuition that,
the back-end has very little computing overhead 
and it can be made reliable due to the simplicity.

\subsection{Effects of Batching and Caching}


While batching and caching are applicable to queue and stack,
only a little cache is needed in both queue and stack to reach a high performance as shown in Table~\ref{tab:opt}.
Also, the performance is less sensitive to the batch size. 
Thus, we do not discuss them here.

\noindent\textbf{Batching.}
We measure the performance of batching with 
vector operations under different batch sizes from 1 to 4048.
The results are in  Figure~\ref{fig:batch-size}. 
MV-BST can be improved by 3.38$\times$ (from 17.8 KOPS without batch to 60.2 KOPS with batch size 1024). 
The improvement for MV-BPT is about 3.13$\times$ (from 28.4 KOPS to 88.9 KOPS). 
The improvements are 126\%, 139\%, and 63\% for BST, BPT, and SkipList, respectively. Multi-version data structures need to 
perform path copying which incurs many write operations. 
The batching can effectively reduce such overhead.

\noindent\textbf{Caching.}
We measure the benefit of caching under different front-end cache sizes. Binary search tree, B+Tree, hashtable, and skip-list are used here, and the results are shown in Figure~\ref{fig:cache-size}. Overall, the throughput increases with the increase of cache sizes. Notice that MV-BPT and MV-BST do not get too much improvement with catching. 
This is due to the fact that the data modified are still kept in
memory for multi-version data structures.
We also measure the improvement due to our special optimizations in the tree-like data structures. The results show that, when using native LRU strategy (i.e., access any data including the lower level nodes through the front-end cache), the BPT can only reach 42.5 KOPS which is 44 \% lower than \anvm.

\subsection{Multiple Front-end Nodes}
The results so far are based on
one front-end and one back-end node. 
We measure the scalability of \anvm 
using multiple readers. 
The results are shown in Figure~\ref{fig:scale-up}. 
We choose five data structures which we mentioned in Section~\ref{sec:concurrency} to make comparisons between lock based and lock-free data structures. 

The readers' performance can scale well with the increasing 
number of front-end nodes. 
We see that, the writer performance of lock based data structure decreases more than that of multi-version data structures. This is because there are more RDMA rounds for lock based data structures that can influence the performance.
With different mechanisms of concurrent control, the effects are different. With lock based BST, the average throughput with 6 readers performs 26\% worse than the value with only one reader. In the case of MV-BST,  performance degradation is about 8\%. 
The results confirm that the multi-version data structures do benefit the readers. 


We also see that, the lock-free data structures scale better than their lock-based counterparts. The readers in Figure~\ref{fig:scale-up-b} have about 3.0$\sim$3.2 $\times$ higher performance than the readers in Figure ~\ref{fig:scale-up-a}. 
Retries incurred by the failed read is the main cause 
for the lower performance. The portion of retry is about 4\%$\sim$16\% of total operations with 6 readers and 100\% insert from the writer. Lower write workload will decrease the ratio of retries. 




We also measure the throughput of multiple front-end 
nodes sharing one NVM device, each accessing its own distinct data structure. 
We only test the case that each front-end use the same type of data structure but with different instances.  Figure~\ref{fig:scale-out} shows that
the scalability is almost linear with 7 front-ends. The average performance degradation for a single client is about 7\% $\sim$ 20\% compared to the one-to-one deployment.

\subsection{Multiple Back-end Nodes}

\begin{figure}[h]
 \vspace{-1em}
  \centerline{\includegraphics[width=0.8\columnwidth]{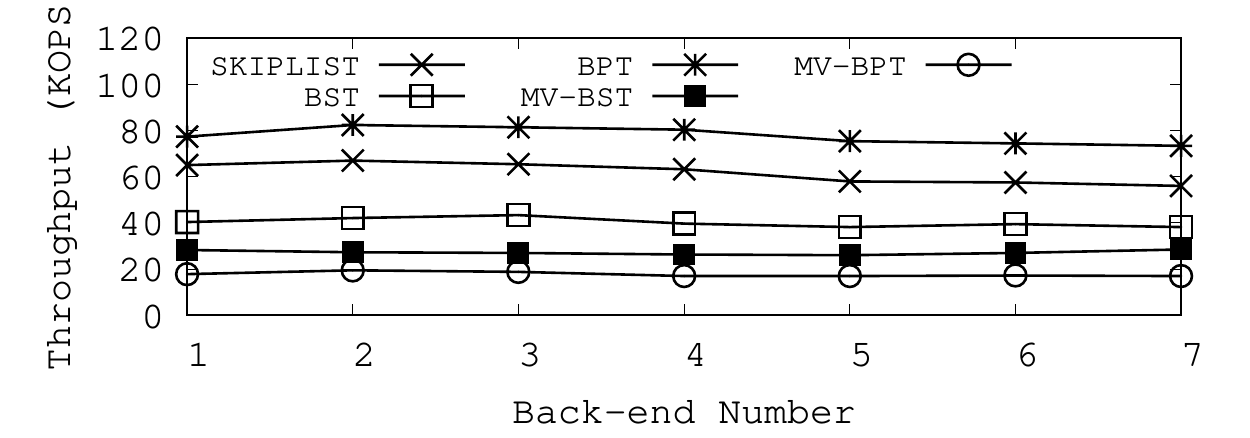}}
\caption{Throughput with Partitions}
\label{fig:back-end}
\end{figure}

As shown in Figure~\ref{fig:back-end}, we measure the performance after partitioning data structure to multiple back-ends. The results show no significant performance degradation after partitioning. The reason is that the partition in each back-end is strictly isolated with other back-ends.

\subsection{Industry Workloads}
We also measure our data structure implementations 
under industry workloads from an online service. The workloads trace the real world user behaviors and satisfy the power-law distribution. 

We also use the operation traces of industry workloads
from online service to evaluate \anvm.
Figure~\ref{fig:workload} shows the throughput of using different read/write ratios from a single writer front-end node. 
For simplicity, insert operation is used as write and find operation is used as read. 
With fewer read operations, the performance decreases 
due to more overhead brought by write operations.
Comparing BPT/BST to their MV-counterparts (MV-BPT/MV-BST), BPT/BST have relatively higher performance. For example, with the full write workload, there are about 23\% and 48\%  performance gap. This is because, in the MV-version, the write operations need to write more data during path copying. 

\section{Related Work}\label{sec:rel}
Single-node NVM systems~\cite{qureshi2009scalable, yang2015nv, kolli2016high, xia2017hikv, chatzistergiou2015rewind, giles2015softwrap, liu2017dudetm, coburn2011nv, korgaonkar2018density, kateja2017viyojit} provide direct access to NVM via memory bus but cause lower utilization of NVM and inaccessible facing node failures.
Distributed NVM systems including 
Octopus~\cite{lu2017octopus}, Hotpot~\cite{shan2017distributed}, Mojim~\cite{zhang2015mojim}, and FaRM~\cite{dragojevic2014farm, dragojevic2015no} combine the NVM devices together with RDMA, and they are all using symmetric deployment. 
Currently, the asymmetric deployments provide storage interfaces including NVMe over Fabric~\cite{couvert2016high} and Crail~\cite{stuedi2017crail}. However, they are not byte addressable i.e., they cannot provide data structure level service.
A few file systems (e.g., Aerie~\cite{volos2014aerie}) adopt a hybrid paradigm like \anvm which allowing direct remote read and transactional write with logs. Different from Aerie, \anvm is a distributed system. 



Recent works on implementation of
a persistent allocation system over NVM include nvm\_malloc~\cite{schwalb2015nvm}, Makalu~\cite{bhandari2016makalu}, PAllocator~\cite{oukid2017memory}, Mneosyne~\cite{volos2011mnemosyne}. 
They discuss considerations of NVM allocators in a single machine. We make the first step towards distributed NVM allocator. 

Several projects aim to design the future disaggregation data center, like~\cite{han2013network, seamicro, intelrackdc, intelrsddc, hpdc, legoos, klimovic2016flash, nanavati2017decibel}. LegoOS~~\cite{legoos} proposes splitkernel, an OS model disseminates functionalities into loosely-coupled monitors. Some of these works focus on how to design remote memory. Aguilera {\em et al.}~\cite{aguilera2017remote} introduce benefits and challenges about applying remote memory. NAM-BD~\cite{zamanian2017end} proposes Network-Attach-Memory (NAM) architecture and implements a time-stamp based concurrency control algorithm. Distinct from other works, INFINISWAP~\cite{infiniswap} provides a page mapping mechanism for memory disaggregation and a decentralized resource management. \anvm is an asymmetric architecture that can be 
used to organize the disaggregated NVM resource. 

\section{Conclusion}\label{sec:con}
This paper rethinks NVM deployment and 
makes a case for the {\em asymmetric} non-volatile
memory architecture, 
which decouples servers from persistent data storage.
We build {\em \anvm framework} based on 
\anvm architecture that implements:
1) high performance persistent data structure update;
2) NVM data management;
3) concurrency control; and 
4) crash-consistency and replication. 
The central idea is to use {\em operation logs}
to reduce the stall due to RDMA writes and 
enable efficient batching and caching in front-end
nodes. 
In a cluster with ten machines (at most seven machines
to emulate a 60GB NVM using DRAM with additional latency),
the results show that \anvm achieves comparable
(sometimes better)
performance to the best possible symmetric architecture while avoiding
all the drawbacks with disaggregation.
Compared to the baseline \anvm, speedup brought by the proposed optimizations is drastic, 
--- 6$\sim$22$\times$ among all benchmarks. 

\end{document}